\documentclass[10pt, conference, compsocconf]{IEEEtran}
\usepackage[sort,compress]{cite}
\usepackage[margin=0pt,labelsep=space,scriptsize,labelfont=bf,textfont=it,up,belowskip=-8pt,aboveskip=5pt]{caption}
\usepackage{graphics,epsfig,graphicx,float,subfigure,color}
\usepackage{algorithmic}

\usepackage[section]{algorithm}

\usepackage{etex}

\usepackage{amsmath,amssymb,amsbsy,amsfonts,latexsym}
\usepackage{multicol,multirow}
\usepackage{hhline}
\usepackage{paralist}
\usepackage{array}
\usepackage{marvosym}
\usepackage{url}
\usepackage{boxedminipage}
\usepackage{wrapfig}
\usepackage{multirow}
\usepackage{threeparttable}
\usepackage{pdfpages}
\usepackage{xspace}
\usepackage{colortbl}

\usepackage{tabularx,booktabs}

\usepackage[usenames,dvipsnames]{xcolor}

\usepackage{tikz} 
\usetikzlibrary{arrows} 
\usetikzlibrary{patterns}  
\usepackage{pgfplots} 
\pgfplotsset{compat=1.3}

\let\labelindent\relax
\usepackage{enumitem}

\usepackage{calligra}
\DeclareMathAlphabet{\mathcalligra}{T1}{calligra}{m}{n}

\def\gap{\hspace*{.2in}}

\newcommand{\figref}[1]{Figure~\ref{#1}}
\newcommand{\tabref}[1]{Table~\ref{#1}}
\newcommand{\secref}[1]{\S\ref{#1}}
\newcommand{\algref}[1]{Algorithm~\ref{#1}}

\renewcommand{\b}[1]{{#1}}

\newcommand{\MA}[1]{{\mathcal #1}}
\DeclareMathOperator{\bigO}{\mathcal{O}}

\definecolor{light-gray}{gray}{0.80}

\newcounter{magicrownumbers}
\newcommand\rownumber{\refstepcounter{magicrownumbers}\arabic{magicrownumbers}}

\newcommand{\bu}{{\b u}} 
\newcommand{\bw}{{\b w}} 
\newcommand{\bx}{{\b x}} 
\newcommand{\Ker}{\mathcal{K}}

\newcommand{\XX}{{\MA{X}}} 
\newcommand{\XXa}{{\MA{X}_\alpha}} 
\newcommand{\sk}[1]{{\widetilde{#1}}} 
\newcommand{\lc}{{\tt l}}  
\newcommand{\rc}{{\tt r}}  
\newcommand{\cc}{{\tt c}}  
\newcommand{\ns}{{s}} 

\newcommand{\ppl}{{m}}
\newcommand{\level}{{l}}
\newcommand{\depth}{{\mathcal{D}}}

\newcommand{\idtol}{{\tau}}
\newcommand{\levrest}{{L}}

\newcommand{\nsmax}{{s_{\textrm{max}}}}

\newcommand{\ASKIT}{{\sc ASKIT}}
\newcommand{\GMRES}{{\sc GMRES}}

\newcommand{\nk}{{\kappa}} 

\renewcommand{\baselinestretch}{0.96}

\begin{document}

\title{An $N \log N$ Parallel Fast Direct Solver for Kernel Matrices}

\author{
\IEEEauthorblockN{Chenhan D. Yu\IEEEauthorrefmark{1}, William B. March\IEEEauthorrefmark{2} and George Biros\IEEEauthorrefmark{4}}
\IEEEauthorblockA{
\IEEEauthorrefmark{1}Department of Computer Science\\
\IEEEauthorrefmark{1}\IEEEauthorrefmark{2}\IEEEauthorrefmark{3}
Institute for Computational Engineering and Science\\
The University of Texas at Austin, Austin, Texas, USA \\
\IEEEauthorrefmark{1}chenhan@cs.utexas.edu,
\IEEEauthorrefmark{2}march@ices.utexas.edu,
\IEEEauthorrefmark{2}gbiros@acm.org
}
}

\maketitle

\begin{abstract}
Kernel matrices appear in machine learning and non-parametric
statistics.  Given $N$ points in $d$ dimensions and a kernel function
that requires $\bigO(d)$ work to evaluate, we present an $\bigO(d
N \log N)$-work algorithm for the approximate factorization of a
regularized kernel matrix, a common computational bottleneck in the
training phase of a learning task.  With this factorization, solving a
linear system with a kernel matrix can be done with $\bigO(N\log N)$
work. Our algorithm only requires kernel evaluations and
does \emph{not} require that the kernel matrix admits an efficient
global low rank approximation.  Instead our factorization only assumes
low-rank properties for the \emph{off-diagonal blocks} under an
appropriate row and column ordering. We also present a hybrid method
that, when the factorization is prohibitively expensive, combines a
partial factorization with iterative methods. As a highlight, we are
able to approximately factorize a dense $11M\times11M$ kernel matrix
in 2 minutes on 3,072 x86 ``Haswell'' cores and a $4.5M\times4.5M$
matrix in 1 minute using 4,352 ``Knights Landing'' cores.
                   

\end{abstract} 

\section{Introduction} \label{s:intro} Let $\mathcal{X}$ be a set of $N$ points $\in \mathbb{R}^{d}$ and let 
$\Ker(\bx_i, \bx_j):\mathbb{R}^d\times\mathbb{R}^d\rightarrow \mathbb{R}$ be a 
given kernel function. The \emph{kernel matrix} is the $N\times N$ matrix whose
entries are given by $K_{ij} = \Ker(\bx_i,\bx_j)$ for
$i,j=1,\ldots,N$, $\bx_i, \bx_j,\in \mathcal{X}$.  

Kernel matrices appear in unsupervised and supervised statistical
learning, Gaussian process, regression, and non-parametric
statistics~\cite{gray-moore01,rasmussen-williams06,
wasserman04,hofmann-scholkopf-smola08}. Solving linear systems with
kernel matrices is an algebraic operation that is required in many
kernel methods. The simplest example is ridge regression in which we
solve 
$\lambda I + K$, where $\lambda>0$
is a regularization parameter that controls generalization accuracy
and $I$ is the identity matrix.  This linear solve can be
prohibitively expensive for large $N$ because $K$ is typically \textbf{\textit{dense}}.
For example, consider the Gaussian kernel,
\begin{equation}\label{e:gaussian}
\Ker(\bx_i,\bx_j) = \exp\left(-\frac{1}{2} \frac{\|\bx_i-\bx_j\|_2^2}{h^2}\right),
\end{equation}
where $h$ is the {\em kernel bandwidth}.  For small $h$, $K$ approaches the
identity matrix whereas for large $h$, $K$ approaches the rank-one
constant matrix. The first regime suggests sparse approximations
while the second regime suggests global low-rank approximations. But
for the majority of $h$ values, $K$ is neither sparse nor globally
low-rank.  Direct factorization of $\lambda I + K$ requires
$\bigO(N^3)$ work, whereas a Krylov iterative method costs
$\bigO(N^2)$ work per iteration and may require 1000s of
iterations. This {\em complexity barrier} has limited the use of kernel
methods for large-scale problems~\cite{lu-may-e15,dai-song-balcan14}. 

{\bf Contributions.} We exploit \emph{hierarchically low-rank}
approximations in which we assume that $K$ can be approximated well by
$D+UV$, where $D$ is block-diagonal and the $U$ and $V$ matrices have
low rank. Using such a decomposition, we improve the factorization
algorithm presented in \cite{yu-march-xiao-biros16}. In that paper the
block-diagonal plus sparse decomposition was done using \ASKIT{}%
\footnote{\scriptsize{
\emph{A}pproximate \emph{S}keletonization \emph{K}ernel \emph{I}ndependent
\emph{T}reecode. The \ASKIT{} library is available
at~\url{http://padas.ices.utexas.edu/libaskit}.}}, a method introduced
in~\cite{march-xiao-biros15,march-xiao-biros-fmm-e15}
(see \S\ref{s:hier}).  \ASKIT{} approximates $K$ in $\bigO( d N\log
N)$ time and the algorithm in~\cite{yu-march-xiao-biros16} factorizes
the \ASKIT{} approximation in $\bigO(N \log^2 N)$ time
(see \S\ref{s:seq}).  Roughly speaking, \ASKIT{} is based on the
approximation of $K$ as the sum of a block-diagonal matrix and a
low-rank matrix followed by recursion for each diagonal block. We
refer this process as the construction of the \emph{hierarchical
representation} of $K$.  Once we have this representation, we can
factorize $K$ by applying recursively the Sherman-Morrison-Woodbury
(SMW) formula. The factorization has to be done for different values
of $\lambda$ during cross-validation studies. Therefore optimizing the
factorization is crucial for the overall performance of a kernel
method. In this paper, we extend the factorization scheme presented
in~\cite{yu-march-xiao-biros16} in several ways:
\begin{itemize}
\item We present an algorithm that factorizes the \ASKIT{} approximation in
$\bigO(N\log N)$ time instead of $\bigO(N\log^2 N)$ and we demonstrate
its performance on several datasets (see \S\ref{s:iter}
and \S\ref{s:results}).
\item We present a hybrid {\em level-restricted} factorization
scheme that reduces dramatically the factorization time by using a
Krylov iterative solver on a much smaller system than the original
(see \S\ref{s:iter}). The new method can be used with matrices for
which \cite{yu-march-xiao-biros16} fails.
\item We study performance on Intel's ``Knights Landing'' (KNL) architecture. 
We introduce an optimized matrix-free kernel summation that reduces
the storage requirements of the factorization without having a very
significant impact on wall-clock time (see \secref{s:ks}).\\
\end{itemize}
In our numerical experiments, we measure the performance of the method
on several different datasets. \ASKIT{} has been applied to
polynomial, Matern, Laplacian, and Gaussian kernels in arbitrary
dimensions.  Due to space limitations we only present the
factorization results for the Gaussian kernel function. Also the
Gaussian kernel is among the hardest to compress in high
dimensions. We examine the performance of the method for different
bandwidth ranges that are relevant to learning tasks, its sensitivity
to the regularization parameter $\lambda$, its performance as we
increase the number of points, and its numerical stability
(see \secref{s:theory}).
%

{\bf Limitations.} Not all kernel matrices admit a good hierarchical
low-rank decomposition. Typically this is related to the intrinsic
dimensionality of the dataset at different scales. So \ASKIT{} and
subsequently our method can fail. If \ASKIT{} can compress the matrix,
the second potential point of failure is the choice of regularization
parameter. If it is too small, our algorithm (as well as
\cite{yu-march-xiao-biros16}) can become numerically unstable.  We
can numerically detect the instability, but it is not clear how to fix
it while maintaining the log-linear complexity of the
algorithm. However, small regularization often results in poor learning
performance so this corner case is not important in applications. We
discuss this in more detail in~\S\ref{s:theory}
and \S\ref{s:results}. Also our methods cannot be
applied to hierarchical decompositions in which $D$ is sparse and not
just block diagonal. For such decompositions, our method can be used as a
preconditioner, as discussed in~\cite{yu-march-xiao-biros16}.

{\bf Related work.} Nystrom methods and their
variants~\cite{williams-seeger01,gittens-mahoney13,hsieh-si-dhillon14,dhillon-si-hsieh14}
can be used to build fast factorizations. However, not all kernel
matrices can be approximated well by Nystrom
methods~\cite{kondor-teneva14,march-xiao-biros-e15,mahoney-darve15,lee-gray-moore06}.
Factorization methods based on hierarchical decomposition have been
studied for kernel matrices from points in two or three
dimensions~\cite{greengard-neil-e14,ambikasaran-darve13,bebendorf08,hackbusch2008approximate},
but less so in high dimensions with a few
exceptions~\cite{kondor-teneva14,yu-march-xiao-biros16}.  Early works
discussing parallel operations for hierarchical matrices on shared
memory system include bulk synchronous parallelization \cite{32416}
and DAG-based task parallelism \cite{66699}.  Distributed
factorization and operations were discussed
in \cite{wang-li-dehoop13,izadi2012parallel}.  The difficulties of
generalizing low-dimensional factorizations in high-dimensions are
discussed in~\cite{march-xiao-biros15,yu-march-xiao-biros16}.
          

\section{Methods} \label{s:methods} 

We begin with a sketch of hierarchical matrices and direct solvers
in \secref{s:hier}. We also briefly summarize the
\ASKIT{} algorithm which we use as the basis for our new methods.
We describe parallel factorization schemes in \secref{s:seq}
and highlight the novelty of our approach
over \cite{yu-march-xiao-biros16}. We then introduce our hybrid
iterative/direct solver in \secref{s:iter}.



  \subsection{Hierarchical Matrices and Treecodes} \label{s:hier}
  
Broadly speaking, we consider a matrix $K\in \mathbb{R}^{N\times N}$ to be
\emph{hierarchical} if it can be partitioned as
\begin{equation}
  \label{e:partitioning}
  K =
  \begin{bmatrix}
  K_{\lc\lc} & K_{\lc\rc}\\
  K_{\rc\lc} & K_{\rc\rc} \\
  \end{bmatrix} \\
   = 
\begin{bmatrix}
K_{\lc\lc} & 0 \\ 
0 & K_{\rc\rc} \\ 
\end{bmatrix} + 
\begin{bmatrix} 
0 & K_{\lc\rc} \\ 
K_{\rc\lc} & 0 \\ 
\end{bmatrix}.
\end{equation} 
where the \emph{off-diagonal} blocks $K_{\lc\rc}$ and $K_{\rc\lc}$ can
be accurately \emph{approximated} by a low-rank factorization and
the \emph{on-diagonal} blocks $K_{\lc\lc}$ and $K_{\rc\rc}$ are
themselves hierarchical. Note that the low-rank structure is \emph{not
invariant} on permutations, it very strongly depends on the ordering
of the columns (or rows since the matrix is symmetric).
For notational convenience we write $K \approx \sk{K} =
D + UV$, where $U$ and $V$ are rank $\ns$ and $D$ is also
hierarchical, where use $\sk{K}$ to indicate the approximate kernel
matrix.

\textbf{Inverting hierarchical matrices.}
When $K$ admits this hierarchical low-rank approximation, then we can
efficiently approximate $K^{-1}$ using the Sherman-Morrison-Woodbury
formula along with recursion:
\begin{equation}\label{e:smw}
\begin{split}
\sk{K}   &= D+UV = D(I+WV), \mbox{~and~} W=D^{-1}U.\\
\sk{K}^{-1} &= (I+WV)^{-1} D^{-1}\\
      &= (I-W(I+VW)^{-1}V)D^{-1} \\
      &= (I-WZ^{-1}V)D^{-1}, \mbox{~where~} Z=I+VW.
\end{split}
\end{equation}
Recursion is used to invert $D^{-1}$. 
After obtaining $D^{-1}$ we compute $W = D^{-1} U$ for a rank-$\ns$
matrix $U$ and factorize the smaller reduced system $Z=I + V
W \in \mathbb{R}^{\ns \times \ns}$.  The scheme can be easily
extended to invert $\lambda I + K$.

To turn this formulation into an algorithm, we need (1) a method to
partition $K$ so that off-diagonal blocks have low-rank, (2) an
efficient way to compute the low-rank factors $U$ and $V$, and (3) a
scheme to construct the inverse.  For the first two tasks we
use \ASKIT{}, a method we recently developed~\cite{march-xiao-biros15,
march-xiao-biros-fmm-e15, march-xiao-biros-e15}. \ASKIT{} uses
geometric information (the input points) to permute $K$ by
partitioning the points recursively using a binary tree. Interactions
between points in a treenode correspond to diagonal blocks of $K$. In
the recursion, the children of the node can be used to define the
block partitioning of the parent block, similar
to \eqref{e:partitioning}.  Next, we summarize \ASKIT{} features
that are necessary for this paper. Please
see~\cite{march-xiao-biros-e15} for the complete details on \ASKIT{}.


\textbf{Partitioning the matrix.}
We use a ball tree \cite{omohundro1989five} to partition $\XX$.
Starting with the root node (which contains the entire data set),
nodes are partitioned into two children (with an equal number of
points) by a splitting hyperplane. This recursive splitting terminates
when a node has less than $\ppl$ points, a user-specified
parameter. The root has \emph{level} $\level=0$ and the leaves
$\level=\depth=\log_2(N/\ppl)$, the \emph{depth} of the tree.  In the
following, we overload $\alpha$, $\beta$ to indicate both \emph{binary tree
nodes} and the \emph{indices of the points} that belong to these nodes;
$|\alpha|$ is the number of points in $\alpha$; and $\lc$, $\rc$
indicate the left and right children of the node. We define
$\XX \in \mathbb{R}^{d \times N}$ to be the matrix of all points and
$\XX_{\alpha}$ to be the points owned by tree node $\alpha$, (i.e.,
$\XX_{\alpha}=\{ \bx_i \lvert \forall i \in \alpha\}$).


\textbf{Computing low rank approximations.}
Let $\alpha$ be the points in a leaf node.  Let $S=\{1,\ldots,N\}\backslash\alpha$.
\emph{The skeletonization} of a node $\alpha$ is a rank-$\ns$ approximation of $K_{S\alpha}$ using 
$\ns$ columns of $K_{S \alpha}$. We refer to these columns as
the \emph{skeleton} of $\alpha$, denoted by $\sk{\alpha}$.
Skeletonization is done using the Interpolative Decomposition
(ID)~\cite{halko-martinsson-tropp11}. Using a pivoted rank-revealing
QR factorization, the ID finds $\sk{\alpha}$ and
$P_{\sk{\alpha} \alpha} \in \mathbb{R}^{\ns\times|\alpha|}$ such that
\begin{equation}\label{e:id}
  K_{S \alpha} \approx K_{S \sk{\alpha}} P_{\sk{\alpha} \alpha}.
\end{equation}
The first $\ns$ pivots from the QR define $\sk{\alpha}$. Using the QR,
we can compute
$P_{\sk{\alpha} \alpha}=K_{S \sk{\alpha}}^{\dagger}K_{S \alpha}$. This
scheme however results in $\bigO(d N^2\ppl)$ complexity for the
overall factorization. We can turn it to a $\bigO(d \log N \ppl)$
scheme by sampling a small subset $S'$ of $S$ and using it instead of
$S$~\cite{march-xiao-biros-e15}.
The approximation rank $\ns$ is chosen such that
$\sigma_{\ns+1}(K_{S' \alpha}) / \sigma_1(K_{S' \alpha}) < \idtol$,
where $\idtol$ is user-specified and $\sigma$ are the singular
values estimated by the diagonal of the rank-revealing QR.

For a non-leaf $\alpha$, we first compute the skeletons $\sk{\lc}$ and
$\sk{\rc}$ of the children of $\alpha$ and then we compute the skeleton
$\sk{\alpha} \subset \sk{\lc} \cup \sk{\rc} = [\sk{\lc} \sk{\rc}]$ and
the projection matrix $P_{\sk{\alpha} [\sk{\lc} \sk{\rc}] }$ using
another ID decomposition (\algref{a:skeletonize}).
Once the skeletonization of every node (but the root) is computed, we
can compute the $i_\mathrm{th}$ entry of $Kw$ by $K_{i:} w \approx
K_{i\alpha}w(\alpha) + \sum_{\level=0}^\depth K_{i \sk{\beta}}
P_{\sk{\beta}\beta} w(\beta)$, where $i\in\alpha$ and $\beta$ are the
siblings of $\alpha$ and its ancestors.
\begin{algorithm}[!t]
\caption{{} [$\sk{\alpha}, P_{\sk{\alpha} \alpha}$]=\texttt{Skeletonize}($\alpha$)}
\begin{algorithmic}
  \STATE \texttt{\bf if} $\alpha$ is leaf \texttt{\bf then return} [$\sk{\alpha}, P_{\sk{\alpha} \alpha}$] = {\tt ID}$(\alpha)$;
  \STATE $[\sk{\lc},]=\texttt{Skeletonize}(\lc)$; $[\sk{\rc},]=\texttt{Skeletonize}(\rc)$;
  \STATE \texttt{\bf return} [$\sk{\alpha}, P_{\sk{\alpha} [\sk{\lc} \sk{\rc}]
  }$] = {\tt ID}($[\sk{\lc} \sk{\rc}]$);
\end{algorithmic}
\label{a:skeletonize}
\end{algorithm}

In \ASKIT{} and \cite{yu-march-xiao-biros16}, 
$U$ and $V$ of a node $\alpha$ for \eqref{e:smw} are
\begin{equation}\label{e:uv}
  \begin{bmatrix}
  0 & K_{\lc\rc} \\
  K_{\rc\lc} & 0 \\
  \end{bmatrix} \approx
  U_\alpha V_\alpha =
  \begin{bmatrix}
  & K_{\lc\sk{\rc}} \\
  K_{\rc\sk{\lc}} & \\
  \end{bmatrix}
  \begin{bmatrix}
  P_{\sk{\lc}\lc} \\
  & P_{\sk{\rc}\rc} \\
  \end{bmatrix}.
\end{equation}
In this work we use the fact that $K_{\lc\rc} = K_{\rc\lc}^{T}$. 
Thus, $K_{\lc\rc}$ can be approximate in two equivalent forms:
$K_{\lc\sk{\rc}}P_{\sk{\rc}\rc}$ or $P_{\lc\sk{\lc}}K_{\sk{\lc}\rc}$.
Here we use the second form to write
\begin{equation}\label{e:uv}
  \begin{bmatrix}
  0 & K_{\lc\rc} \\
  K_{\rc\lc} & 0 \\
  \end{bmatrix} \approx
  U_\alpha V_\alpha =
  \begin{bmatrix}
  {P}_{\lc\sk{\lc}} & \\
  & {P}_{\rc\sk{\rc}} \\
  \end{bmatrix}
  \begin{bmatrix}
  & K_{\sk{\lc}\rc} \\
  K_{\sk{\rc}\lc} & \\
  \end{bmatrix}.
\end{equation}
We will see that having the $P$ terms on the left allows
us to design an $\MA{O}(N\log N)$ factorization algorithm.


\textbf{Level restriction.}
We provide details on the level-restriction feature of \ASKIT{}.  As
we saw in~\algref{a:skeletonize}, the skeletonization proceeds in a
bottom-up traversal of the ball tree.  As we traverse the tree, the
off-diagonal blocks are growing larger and, depending on the problem,
the necessary rank $s$ can increase to the extend that no compression
takes place. To guarantee accuracy, skeletonization of
$\alpha$ should terminate if $\sk{\alpha}=\sk{\lc}\cup\sk{\rc}$.  In
some of our numerical experiments, instead of using this criterion, we
use $\levrest$ to represent the level at which the skeletonization
stops.

With level restriction, the factorization described
in~\cite{yu-march-xiao-biros16} \emph{cannot be used}.  We introduce a
hybrid iterative/direct scheme that addresses this shortcoming
in \secref{s:iter}.


  \subsection{Fast direct solver} \label{s:seq} We have sketched how we compute the $UV$ approximations in $\sk{K}$
using \ASKIT{}.  We now discuss how to use them in the context
of \eqref{e:smw} to solve $\lambda I + K$ directly. For simplicity, we
describe the case where $\lambda = 0$, but all the algorithms we
describe trivially generalize to the $\lambda \neq 0$ case. We first
consider the case in which no level restriction takes place.

We assume that all the internal nodes have been skeletonized. The
factorization of $\sk{K}$ proceeds using a bottom-up traversal of
the tree.  At the leaf level, we factorize
$K_{\alpha \alpha}^{-1} \in \mathbb{R}^{\ppl \times \ppl}$ using
LAPACK's \texttt{GETRF}.  For an internal node $\alpha$, we need $U_{\alpha}$,
$W_{\alpha}$, $V_{\alpha}$, and $Z^{-1}_{\alpha}$.
Using \eqref{e:smw}, we can write out
$\sk{K}_{\alpha\alpha} = D_{\alpha} (I + W_{\alpha}V_{\alpha})$
as
\begin{equation}
\label{e:hodlr}  
  \begin{bmatrix}
  \sk{K}_{\lc\lc} &  \\
  & \sk{K}_{\rc\rc} \\
  \end{bmatrix}
  \left(
  I
  +
  \begin{bmatrix}
  \hat{P}_{\lc\sk{\lc}} & \\
  & \hat{P}_{\rc\sk{\rc}} \\
  \end{bmatrix}
  \begin{bmatrix}
  & K_{\sk{\lc}\rc} \\
  K_{\sk{\rc}\lc} & \\
  \end{bmatrix}
  \right),
\end{equation}
where we define
$\hat{P}_{\lc\sk{\lc}}=\sk{K}_{\lc\lc}^{-1}P_{\lc\sk{\lc}}$ and
$\hat{P}_{\rc\sk{\rc}}=\sk{K}_{\rc\rc}^{-1}P_{\rc\sk{\rc}}$ (notice
the ``hat'' notation).  Therefore, $\hat{P}_{\alpha\sk{\alpha}}$
requires ``inverting'' $\sk{K}_{\alpha\alpha}^{-1}$, which in turn
requires traversing all the descendants of $\alpha$ (the subtree
rooted at $\alpha$) and recursively applying \eqref{e:hodlr}. (We
introduced this scheme in~\cite{yu-march-xiao-biros16} and results in
$\bigO(d N \log^2 N)$ complexity.)  But as we will see shortly this
subtree traversal is not necessary.)
Once we have $W_{\alpha}$, we use the SMW formula to invert $(I+W_{\alpha}V_{\alpha})^{-1}$. This inverse requires  $W_{\alpha}Z^{-1}_{\alpha}V_{\alpha}$, which in terms of the block decomposition of $\alpha$, can be written as
\begin{equation}
\label{e:blocksmw}
  \begin{bmatrix}
  \hat{P}_{\lc\sk{\lc}} & \\
  & \hat{P}_{\rc\sk{\rc}} \\
  \end{bmatrix}
  \begin{bmatrix}
  I & K_{\sk{\lc}\rc}\hat{P}_{\rc\sk{\rc}} \\ 
  K_{\sk{\rc}\lc}\hat{P}_{\lc\sk{\lc}} & I \\
  \end{bmatrix}^{-1}
  \begin{bmatrix}
  & K_{\sk{\lc}\rc} \\
  K_{\sk{\rc}\lc} & \\
  \end{bmatrix}.
\end{equation}
Since $\alpha$ is the parent of $\lc$ and $\rc$, 
$\hat{P}_{\lc\sk{\lc}}$ and $\hat{P}_{\rc\sk{\rc}}$ have been already computed. 
$K_{\sk{\lc}\rc}\hat{P}_{\rc\sk{\rc}}$ and $K_{\sk{\rc}\lc}\hat{P}_{\lc\sk{\lc}}$ are computed by 
\texttt{GEMM}, and the reduced system is factorized by \texttt{GETRF}.

We can exploit
a \emph{``telescoping''} relation between $\hat{P}_{\lc\sk{\lc}}$,
$\hat{P}_{\rc\sk{\rc}}$ and $\hat{P}_{\alpha\sk{\alpha}}$.
We say that  $P_{\alpha\sk{\alpha}}$ is ``telescoped''  from
$P_{[\sk{\lc}\sk{\rc}] \sk{\alpha}}$, $P_{\lc \sk{\lc}}$, and 
$P_{\rc \sk{\rc}}$ because is computed by  formula in the box below.
\begin{equation}
  \label{e:innerid}
  D_{\alpha}^{-1}\framebox{$P_{\alpha\sk{\alpha}}$} = 
  \begin{bmatrix}
  \sk{K}_{\lc\lc}^{-1} & \\
  & \sk{K}_{\rc\rc}^{-1} \\
  \end{bmatrix}
  \framebox{$
  \begin{bmatrix}
  P_{\lc\sk{\lc}} &  \\
  & P_{\rc\sk{\rc}} \\
  \end{bmatrix}
  P_{[\sk{\lc}\sk{\rc}]\sk{\alpha}}
  $}.
\end{equation}
The calculation in the box requires just \texttt{GEMM} operations
from the children ($\lc$ and $\rc$) but not all descendants. Since
$\hat{P}_{\alpha\sk{\alpha}}
= \sk{K}_{\alpha\alpha}^{-1}P_{\alpha\sk{\alpha}} =
(I+W_{\alpha}V_{\alpha})^{-1}D_{\alpha}^{-1}P_{\alpha\sk{\alpha}}$,
we can replace $D_{\alpha}^{-1}P_{\alpha\sk{\alpha}}$ with \eqref{e:innerid}.
Now we find that $\hat{P}_{\alpha\sk{\alpha}}$ can also be telescoped by
$\hat{P}_{\lc\sk{\lc}}$ and $\hat{P}_{\rc\sk{\rc}}$ as
\begin{equation} \label{e:invtelescope}
\begin{split}
  \hat{P}_{\alpha\sk{\alpha}} & = 
  \left(I+
  \begin{bmatrix}
  \hat{P}_{\lc\sk{\lc}} & \\
  & \hat{P}_{\rc\sk{\rc}} \\
  \end{bmatrix}
   V_{\alpha} \right)^{-1}
  \begin{bmatrix}
  \hat{P}_{\lc\sk{\lc}} & \\
  & \hat{P}_{\rc\sk{\rc}} \\
  \end{bmatrix}
  P_{[\sk{\lc}\sk{\rc}]\sk{\alpha}}.
\end{split}
\end{equation}
Notice that we no longer need to solve 
$\sk{K}^{-1}_{\lc\lc}$ and 
$\sk{K}^{-1}_{\rc\rc}$ in \eqref{e:invtelescope}.
Thus, no tree traversal is required.
In the leaf level (base case), $\hat{P}_{\alpha\sk{\alpha}}$ is computed
directly from $K_{\alpha\alpha}^{-1}P_{\alpha\sk{\alpha}}$.


Given these formulas and the skeletonization computed in
\algref{a:skeletonize}, we compute the factors needed for the
direct solver in a postorder traversal of the tree
(\algref{a:factor}). If $\alpha$ is a leaf node, we factorize $\lambda
I + K_{\alpha\alpha}$ using an LU factorization.  Otherwise, we
compute $K_{\sk{\lc}\rc}$ and $K_{\sk{\rc}\lc}$.  Notice that
$\hat{P}_{\lc\sk{\lc}}$ and $\hat{P}_{\rc\sk{\rc}}$ are computed in
the previous recursion; thus, we can form and factorize the reduced
system $Z_{\alpha}$.  Finally, $\hat{P}_{\alpha\sk{\alpha}}$ is
telescoped using \eqref{e:invtelescope}, thus
$\texttt{Solve}(\alpha,W_{\alpha}P_{[\sk{\lc}\sk{\rc}]\sk{\alpha}},\texttt{false})$
(\algref{a:solve}) will not invoke recursion.


%
%
%
This algorithm improves on the one in \cite{yu-march-xiao-biros16} by
 removing the extra subtree traversals that result in $\bigO(N\log^2
 N)$ complexity.  Instead, our algorithm exploits the nested structure
 of $\hat{P}_{\alpha\sk{\alpha}}$ resulting in an $N \log N$
 complexity for the factorization. In some of our largest runs, this
 resulted in over 3$\times$ speedup without any change in the
 accuracy.

\begin{algorithm}[!t]
\begin{algorithmic}
  \STATE \texttt{\bf if} $\alpha$ is leaf \texttt{\bf then} 
  \STATE \gap LU factorization $\lambda I + K_{\alpha\alpha}$.
  \STATE \gap $W_{\alpha}=P_{\alpha\sk{\alpha}}$ and $P_{[\sk{\lc}\sk{\rc}]\sk{\alpha}} = I$.
  \STATE \texttt{\bf else}
    \STATE \gap \texttt{Factorize}($\lc$) and \texttt{Factorize}($\rc$). 
    \STATE \gap Form $W_{\alpha}$ with $\hat{P}_{\lc\sk{\lc}}$,
    $\hat{P}_{\rc\sk{\rc}}$, and $V_{\alpha}$ with $K_{\sk{\lc}\rc}$, $K_{\sk{\rc}\lc}$.
    \STATE \gap LU factorize the reduced system $Z_{\alpha}$ in \eqref{e:blocksmw}.
  \STATE $\hat{P}_{\alpha\sk{\alpha}} =
  \texttt{Solve}(\alpha,W_{\alpha}P_{[\sk{\lc}\sk{\rc}]\sk{\alpha}},\texttt{false})$ using \eqref{e:invtelescope}.
\end{algorithmic}
\caption{{} \texttt{Factorize}($\alpha$)}
\label{a:factor}
\end{algorithm}
\begin{algorithm}[!t]
\begin{algorithmic}
  \STATE \texttt{\bf if} $\alpha$ is leaf \texttt{\bf then} LU solver $\bw = (\lambda I + K_{\alpha\alpha})^{-1}\bu$ 
  \STATE \texttt{\bf else}
    \STATE \gap \texttt{\bf if} \texttt{do\_recur} \texttt{\bf then}
    \STATE \gap \gap $v=[\texttt{Solve}(\lc, \bu_{\lc}, \texttt{true})$;
    $\texttt{Solve}(\rc, \bu_{\rc}, \texttt{true})]$. 
    \STATE \gap Compute $\bw=u-W_{\alpha}Z^{-1}_{\alpha}V_{\alpha}u$ using \eqref{e:blocksmw}.
\end{algorithmic}
\caption{{} $\bw=\texttt{Solve}(\alpha,\bu,\texttt{do\_recur})$} 
\label{a:solve}
\end{algorithm} 

We then describe how to apply $\sk{K}_{\alpha\alpha}^{-1}$ to a vector $\bu$,
shown in \algref{a:solve}.
If $\alpha$ is a leaf node, we can directly invoke an LU solver to obtain 
$\bw = K_{\alpha\alpha}^{-1}\bu$. Otherwise, we have two situations.
If \algref{a:solve} is called by \texttt{Factorize} (\texttt{do\_recur} is
\texttt{false}), then we know
$u=W_{\alpha}P_{[\sk{\lc}\sk{\rc}]\sk{\alpha}}$. Thus, no recursion is
required. While \texttt{Solve} is called to solve a random $u$, then we need to
solve
$\sk{K}_{\lc\lc}^{-1}\bu_{\lc}$ and
$\sk{K}_{\rc\rc}^{-1}\bu_{\rc}$ recursively and compute 
\eqref{e:blocksmw} with \texttt{GEMV} on $V_\alpha$ and $W_{\alpha}$ and an LU
solve \texttt{GETRS} on $Z_{\alpha}$.

Therefore, the complete algorithm consists of constructing the tree,
calling \algref{a:skeletonize}, then \algref{a:factor} and \algref{a:solve},
each called on the root of the tree.  


\begin{figure}[!t]
  \centering
  \includegraphics[scale=.3]{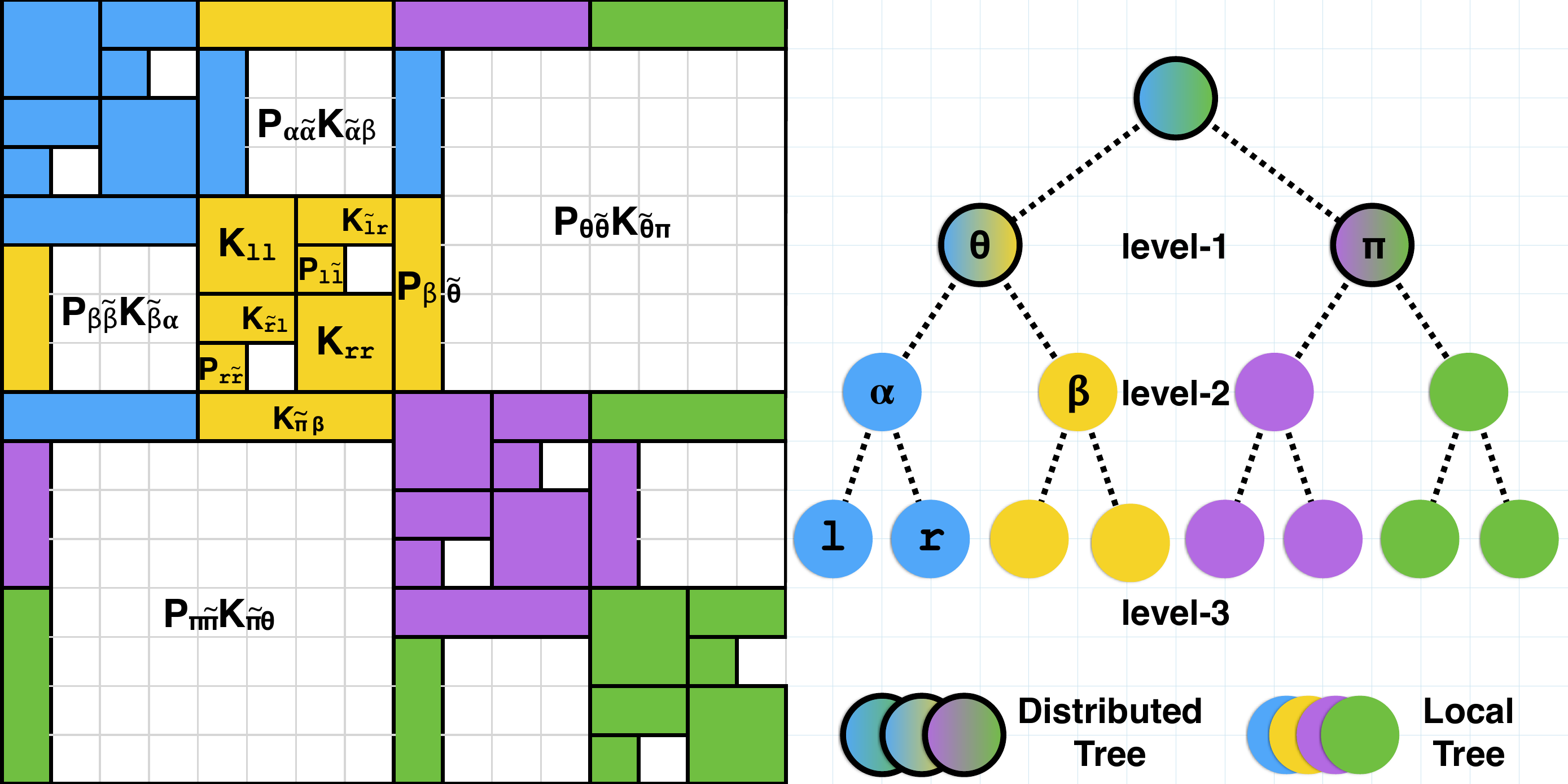}
  \caption{The top four levels of the tree and the corresponding blocks of the 
  matrix $\tilde{K}$. The nodes belonging to each process are highlighted in a 
  single color. Each process factorizes its own portion of the tree independently.
  We also highlight the factors used in the direct solver construction and 
  show which process owns which factor. Each process own a diagonal block and 
  all factors in the same column and the same row. For example, the yellow
process owns $P_{\beta\sk{\beta}}$ and $K_{\sk{\alpha}\beta}$ at level 1;
similarly it owns $P_{\beta\sk{\theta}}$ and $K_{\sk{\pi}\beta}$ at level 0.} 
  \label{fig:distributed_phss}
\end{figure}

\textbf{Parallel direct solver.} The parallelization is essentially
identical to the scheme proposed in~\cite{yu-march-xiao-biros16}. 
Each subtree (a set of points $\{x\}$) is assigned to a distributed-memory process (or a
worker). 
Although we described recursive version of our algorithms we use
level-by-level traversals combined with shared or distributed memory
parallelism (depending on the level) across nodes in the same level.
If the number of nodes is less than the number of physical cores, the
OpenMP nested construct is enabled such that each thread will invoke
parallel BLAS or LAPACK routines.  If we have $p$ processes, then
above level $\log p$ of the tree, we have to communicate to compute
factors, since the terms needed are distributed among processes. We
use the Message Passing Interface (MPI) library for distributed memory
communication.


In \figref{fig:distributed_phss}, we summarize the distributed-memory
algorithm.  The four colors represent four different MPI ranks and the
nodes they own. The tree on the right shows $l=2$ treenodes are
uniquely assigned to ranks, but treenodes with $l>2$ are shared
among ranks.  To facilitate collective communication, each distributed
treenode creates a local communicator, which equally divides the ranks
of the parent.  We use $\{i\}$ to denote the $i_{th}$ MPI rank in the
local communicator.  Consider the communicator of $\alpha$, which
involves $q$ ranks.  Let $\cc$ denote the child of $\alpha$ that
$\{i\}$ owns.  Then $\cc = \lc$ if $\{i<\frac{q}{2}\}$. Otherwise,
$\cc = \rc$.  For a distributed node $\alpha$, data points $\{x\}_i$
owned by $\{i\}$ are never required by other MPI processes, and $\XXa
= \XX_{\lc} \cup \XX_{\rc} = (\cup_{i<\frac{q}{2}}\{x\}_i) \cup
(\cup_{i\geq \frac{q}{2}}\{x\}_i)$.  However, skeletons $\sk{\alpha}$
and $P_{\sk{\alpha}[\sk{\lc}\sk{\rc}]}$ are only stored on
$\{0\}$. When its sibling needs this information, we exchange the
information using a \texttt{SendRecv} between $\{0\}$ and
$\{\frac{q}{2}\}$ using the parent communicator of $\alpha$ and its
sibling. Once received, $\alpha$ and the sibling communicator can
\texttt{Bcast} to every processes in their groups. 


\begin{algorithm}[!t]
\begin{algorithmic}
  \STATE \texttt{\bf if} {$\alpha$ is at level $\log{p}$} \texttt{\bf
  then} \texttt{Factorize($\alpha$)}.  \STATE \texttt{\bf
  else} \STATE \gap \texttt{DistFactorize}($\cc$,
  $\frac{q}{2}$).  \setlength\multicolsep{3.5pt}
\begin{multicols}{2}
  \STATE \gap $\{i<\frac{q}{2}\}$ \hrulefill
  \STATE \gap $\{0\}$\texttt{Send} $\sk{\lc}$.
  \STATE \gap $\{0\}$\texttt{Recv}, \texttt{Bcast} $\sk{\rc}$.
  \STATE \gap \texttt{Reduce} $K_{\sk{\rc}\{x\}}\hat{P}_{\{x\}\sk{\lc}}$.
  \STATE \gap $\{0\}$ \texttt{Recv}
  \STATE $\{i\geq\frac{q}{2}\}$ \hrulefill
  \STATE $\{\frac{q}{2}\}$\texttt{Recv}, \texttt{Bcast} $\sk{\lc}$.
  \STATE $\{\frac{q}{2}\}$\texttt{Send} $\sk{\rc}$.
  \STATE \texttt{Reduce} $K_{\sk{\lc}\{x\}}\hat{P}_{\{x\}\sk{\rc}}$.
  \STATE $\{\frac{q}{2}\}$ \texttt{Send} $K_{\sk{\lc}\{x\}}\hat{P}_{\{x\}\sk{\rc}}$.
\end{multicols}
  \STATE $\{0\}$ \texttt{Bcast} $P_{[\sk{l}\sk{r}]\sk{\alpha}}$ and LU factorizes $Z$.
  \STATE $\hat{P}_{\{x\}\sk{\alpha}} =
  \texttt{DistSolve}(\alpha,\hat{P}_{\{x\}\sk{\cc}}P_{[\sk{\cc}]\sk{\alpha}},q,\texttt{false})$.
\end{algorithmic}
\caption{{} \texttt{DistFactorize}($\alpha$,$q$)}
\label{a:distributedfactor}
\end{algorithm}
\begin{algorithm}[!t]
\begin{algorithmic}
  \STATE \texttt{\bf if} {$\alpha$ is at level $\log{p}$} \texttt{\bf then}
  $w=\texttt{Solve}(\alpha,\bu,\texttt{do\_recur})$. 
  \STATE \texttt{\bf else}
    \STATE \gap \texttt{\bf if} \texttt{do\_recur} \texttt{\bf then}
    \STATE \gap \gap $u=\texttt{DistSolve}(\cc, \bu, \frac{q}{2}, \texttt{do\_recur})$.
\setlength\multicolsep{3.5pt}
\begin{multicols}{2}
  \STATE \gap $\{i<\frac{q}{2}\}$ \hrulefill         
    \STATE \gap \texttt{Reduce} $K_{\sk{\rc}\{x\}}u$.
    \STATE \gap $\{0\}$ \texttt{Recv} $u_{\sk{\lc}}$.
      \STATE $\{i\geq\frac{q}{2}\}$ \hrulefill    
    \STATE \texttt{Reduce} $u_{\sk{\lc}}=K_{\sk{\lc}\{x\}}u$.
    \STATE $\{\frac{q}{2}\}$ \texttt{Send} $u_{\sk{\lc}}$.
\end{multicols}
    \STATE \gap \{0\}  $[u_{\sk{\lc}}; u_{\sk{\rc}}] = Z^{-1}[ u_{\sk{\lc}}; u_{\sk{\rc}}]$ using \eqref{e:blocksmw}.
\begin{multicols}{2}
    \STATE \gap $\{0\}$ \texttt{Send} $u_{\sk{\rc}}$.
    \STATE \gap $\{0\}$ \texttt{Bcast} $u_{\sk{\lc}}$.
    \STATE \gap $w = u - \hat{P}_{\{x\}\sk{\lc}}u_{\sk{\lc}}$.
    \STATE $\{\frac{q}{2}\}$ \texttt{Recv} $u_{\sk{\rc}}$.
    \STATE $\{\frac{q}{2}\}$ \texttt{Bcast} $u_{\sk{\rc}}$.
    \STATE $w = u - \hat{P}_{\{x\}\sk{\rc}}u_{\sk{\rc}}$.
\end{multicols}    
\end{algorithmic}
\caption{{} $\bw=\texttt{DistSolve}(\alpha, \bu, q, \texttt{do\_recur})$}
\label{a:distributedsolve}
\end{algorithm}

\algref{a:distributedfactor} describes the recursive distributed factorization. 
In each node $\alpha$, ranks $\{i<\frac{q}{2}\}$ requires skeletons $\sk{\rc}$
owned by $\{\frac{q}{2}\}$ to compute $K_{\sk{\rc}\{x\}}$, and $\{i\geq\frac{q}{2}\}$ 
requires $\sk{\lc}$ owned by $\{0\}$ to compute $K_{\sk{\lc}\{x\}}$.
Assuming that $\hat{P}_{\{x\}\sk{\lc}}$ and $\hat{P}_{\{x\}\sk{\rc}}$ were
computed, then each rank computes 
$K_{\sk{\rc}\{x\}}\hat{P}_{\{x\}\sk{\lc}}$ and
$K_{\sk{\lc}\{x\}}\hat{P}_{\{x\}\sk{\rc}}$.
$\{0\}$ reduces all $K_{\sk{\rc}\{x\}}\hat{P}_{\{x\}\sk{\lc}}$ to form
$K_{\sk{\rc}\lc}\hat{P}_{\lc\sk{\lc}}$.
$\{\frac{q}{2}\}$ reduces all $K_{\sk{\lc}\{x\}}\hat{P}_{\{x\}\sk{\rc}}$ and
sends $K_{\sk{\lc}\rc}\hat{P}_{\rc\sk{\rc}}$ to $\{0\}$.
The LU factorization is done and stored on $\{0\}$.
Finally, $\{0\}$ needs to broadcast $P_{[\sk{\lc}\sk{\rc}]\sk{\alpha}}$ such
that all processes can telescope $\hat{P}_{\{x\}\sk{\alpha}}$ by the 
distributed solver.

\algref{a:distributedsolve} is the recursive distributed solver, which will be
called in two instances. When it is called by \texttt{DistFactorize}
(\texttt{do\_recur} is \texttt{false}), we know our input $u$ can be
telescoped by \eqref{e:invtelescope} and no recursion takes place.  On
the other hand, if it is called to solve some random $u$, then we need
to traverse all the way down to the leaf level.  Since the column blocks
of $V_{\alpha}$ ($K_{\sk{\lc}\{x\}}$ and $K_{\sk{\rc}\{x\}}$) are
distributed, a MatVec with $V$ requires reduction between processes.
$u_{\sk{\rc}}=K_{\sk{\rc}\{x\}}u$ are reduced by $\{0\}$, and
$u_{\sk{\lc}}=K_{\sk{\rc}\{x\}}u$ are reduced by $\{\frac{q}{2}\}$.
The system $Z^{-1}[u_{\sk{\lc}}; u_{\sk{\rc}}]$ is solved on $\{0\}$.
The output is again broadcast back to all processes, because row
blocks of $W_{\alpha}$ ($\hat{P}_{\{x\}\sk{\lc}}$ and
$\hat{P}_{\{x\}\sk{\rc}}$) are again distributed among processes. Once
reaching level $\log p$,
\algref{a:solve} is called to work on the local subtree.

%
%
%
%


  \subsection{Fast hybrid solver} \label{s:iter} \begin{figure}[!t]
  \centering
  \includegraphics[scale=.3]{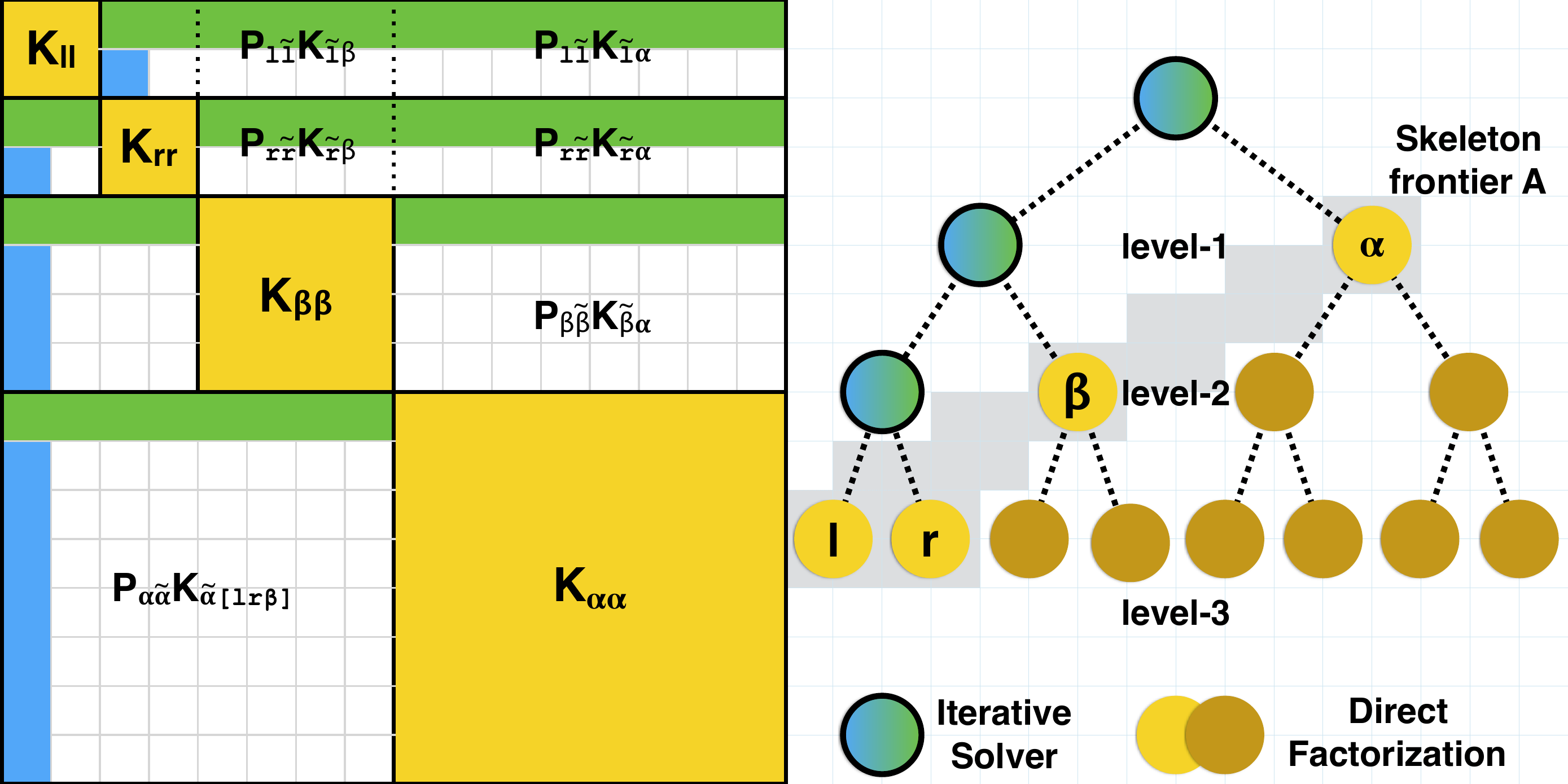}
  \caption{The yellow nodes are the skeletonization frontier $A$.
  Their parents (green and blue nodes) are not skeletonized, because 
	further compression may result in loss of accuracy. We also highlight 
  the expanded matrix blocks which are necessary because of this lack of 
  compression.} 
  \label{fig:frontier}
\end{figure}

As we discussed level-restriction is necessary when an off-diagonal
block is no longer low-rank. We refer to the set of nodes that
skeletonized but whose parent did not as the \emph{skeletonization
  frontier} $A$. In~\figref{fig:frontier}, the yellow nodes define the
frontier. Nodes ``above'' (i.e., closer to the root) the frontier
cannot be skeletonized.

In this case the SMW formulation~\eqref{e:smw} can still be used but
$D$, $U$, and $V$ have as many blocks as the number of nodes above
$A$. Therefore, the $Z$ matrix will be quite large. For example, if the
frontier $A$ consists of all the nodes at $L$, and a node in $A$ has
$\ns$ skeletons, then the size of $Z$ will be $2^L \ns$.  If we
compute the full factorization, the cost will be
$\MA{O}(2^{2L}s^2N+2^{3L}s^3)$ in work and $\MA{O}(2^LsN)$ in storage.
The basic idea here is not to store and factorize any $Z$, $W$ and $V$
factors for those unskeletonized nodes, but instead use a matrix-free
Krylov method. We refer to this approach as the \emph{``hybrid
  method''}, since we factorize only up to frontier. Level restriction
reduces the system that needs to be solved from $N$ to $2^L \ns$.

\begin{figure}[!t]
  \centering
  \includegraphics[scale=.3]{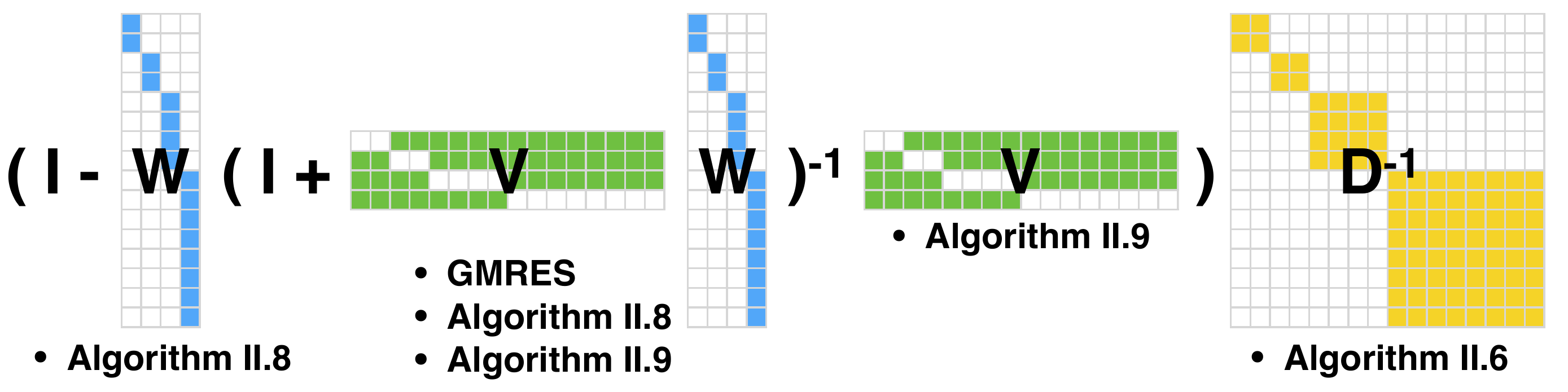}
  \caption{The SMW formula for the partial factorization in
    \figref{fig:frontier}. $D$ is factorized
    by \algref{a:distributedfactor} and solved by
    \algref{a:distributedsolve} on skeletonized nodes. Non-skeletonized nodes
    are collapsed into $W$ and $V$ factors of size $2^L \ns \times N$,
    so they can no longer be factorized efficiently. The
    reduced system $(I+VW)$ is solved iteratively by GMRES.}
  \label{fig:iterative_smw}
\end{figure}
\begin{algorithm}[!t]
\caption{{} $\bw=\texttt{HybridSolve}(\bu,A)$}
\begin{algorithmic}
  \STATE $w(\alpha) = \texttt{DistSolve}(\alpha,\bu,\texttt{true})$ for all $\alpha \in A$.
  \STATE $w=\texttt{MatVecV}(\texttt{root},w,A)$.
  \STATE Solve $w = ( I + VW )^{-1}w$ iteratively.
  \STATE Compute $w = u - \texttt{MatVecW}(\texttt{root},w,A)$.
\end{algorithmic}
\label{a:iterativesolve}
\end{algorithm}
\begin{algorithm}[!t]
\caption{{} $w=\texttt{MatVecW}(\alpha,x,A)$}
\begin{algorithmic}
  \STATE \texttt{\bf if} $\alpha$ is at level-log$p$ \texttt{\bf then}
  \STATE \gap \texttt{\bf if} $\alpha \in A$ \texttt{\bf then} $w=\hat{P}_{\alpha\sk{\alpha}}u$.
  \STATE \gap \texttt{\bf else} $w=[\texttt{MatVecW}(\lc,u_{\lc},A); \texttt{MatVecW}(\rc,u_{\rc},A)]$.
  \STATE \texttt{\bf else}
  \STATE \gap \texttt{\bf if} $\alpha \in A$ \texttt{\bf then} $w=\hat{P}_{\{x\}\sk{\alpha}}u$
  \STATE \gap \texttt{\bf else} $w=\texttt{MatVecW}(\cc,u,A)$.
\end{algorithmic}
\label{a:matvecw}
\end{algorithm}
\begin{algorithm}[!t]
\caption{{} $w=\texttt{MatVecV}(\alpha,y,A)$}
\begin{algorithmic}
  \STATE \texttt{\bf if} $\alpha$ is at level-log$p$ \texttt{\bf then}
  \STATE \gap $w=K_{\sk{\beta}\alpha}u$ ($\beta$ is the sibling of $\alpha$).
  \STATE \gap \texttt{\bf if} $\alpha \notin A$ \texttt{\bf then}
  \STATE \gap \gap $w=[\texttt{MatVecV}(\lc,u_{\lc},A); \texttt{MatVecV}(\rc,u_{\rc},A)]$.
  \STATE \texttt{\bf else}
  \STATE \gap $w=K_{\sk{\beta}\{x\}}u$ ($\beta$ is the sibling of $\alpha$).
  \STATE \gap \texttt{\bf if} $\alpha \notin A$ \texttt{\bf then}
  $w=\texttt{MatVecV}(\cc,u,A)$.
  \STATE \texttt{\bf if} $\alpha$ is \texttt{root} \texttt{\bf then} \texttt{AllReduce} $w$.
\end{algorithmic}
\label{a:matvecv}
\end{algorithm}


\textbf{Partial factorization.} We still factorize skeletonized nodes 
bottom up until reaching the frontier $A$.
In \figref{fig:frontier}, these treenodes (yellow and khaki) we factorize 
are diagonal blocks $D$ (yellow) on the left.
Then, conceptually, we coalesce all $W$ (blue) and $V$ (green) factors for 
nodes above $A$.  
Notice that we can still apply SMW on this
partial factorization in the form of \figref{fig:iterative_smw}.
Rather than continuing to factorize
these unskeletonized nodes, we switch 
to an iterative solution for the reduced system $(I+VW)^{-1}$.

In \algref{a:iterativesolve}, we show this hybrid algorithm.
$D^{-1}$ is computed by \algref{a:distributedsolve} on those 
skeletonized nodes.
The iterative solver requires the ability to compute the MatVec
for $W$ and $V$. 
\algref{a:matvecw} (\texttt{MatVecW}) traverses downward from the root. 
Since $P_{\sk{\alpha}[\sk{\lc}\sk{\rc}]} = I$ for all $\alpha$ 
above the frontier, \texttt{MatVecW} only occurs on the frontier ($\alpha \in A$).
\algref{a:matvecv} (\texttt{MatVecV}) computes $K_{\sk{\beta}\alpha}u$ or 
$K_{\sk{\beta}\{x\}}u$ for all nodes above (including) the frontier.
In the distributed tree, \texttt{MatVecV} performs a reduction on $\{x\}$;
thus, an \texttt{AllReduce} is required at the end such that
all MPI ranks get the same output. On the other hand, \texttt{MatVecW} is
supposed to perform a scattering on $\{x\}$. 
Thus, \texttt{MatVecW} always
happens after \texttt{MatVecV}. In this case, the inputs of \algref{a:matvecw}
for all MPI ranks are the same.

  \subsection{Fast kernel summation} \label{s:ks} \begin{table}[!t]
\setlength\tabcolsep{5.7pt}
\centering
{
  \begin{tabular}{|>{\columncolor[gray]{0.8}}r|r|rrrrrr|} 
  \hline
  \rowcolor[gray]{0.8}
  Arch & $d$ & 4 & 20 & 36 & 68 & 132 & 260 \\
  \hline
  Haswell & \texttt{MKL$+$VML} & 31 & 53 & 72 & 115 & 190 & 305 \\
  16K & \texttt{GSKS} & 321 & 465 &  512 & 634 & 687 & 680 \\
  \hline
  KNL     & \texttt{MKL$+$VML} & 12 & 93 & 132 & 416 & 636 & 916 \\
  16K     & \texttt{GSKS} & 703 & 888 & 1067 & 1246 & 1334 & 1449 \\
  \hline
  Haswell & \texttt{MKL$+$VML} & 32 & 56 & 80 & 110 & 198 & 296 \\
  8K      & \texttt{GSKS} & 301 & 448 &  515 & 558 & 620 & 543 \\
  \hline
  KNL     & \texttt{MKL$+$VML} & 11 & 93 & 103 & 166 & 506 & 753 \\
  8K      & \texttt{GSKS} & 479 & 888 &  903 & 975 & 1220 & 1345 \\
  \hline
  Haswell & \texttt{MKL$+$VML} & 30 & 52 & 70 & 110 & 180 & 284 \\
  4K      & \texttt{GSKS} & 250 & 359 &  384 & 420 & 477 & 468 \\
  \hline
  KNL     & \texttt{MKL$+$VML} & 11 & 56 & 76 & 116 & 370 & 578 \\
  4K      & \texttt{GSKS} & 341 & 445 & 464 & 510 & 858 & 1015 \\
  \hline
  \end{tabular}
}
\caption{Gaussian kernel summation efficiency of 
  $16K\times16K\times d$,
  $8K\times8K\times d$,
  and $4K\times4K\times d$ in \rm{GFLOPS}. 
  \texttt{GSKS} can be found in https://github.com/ChenhanYu/ks.
  The reference implementation uses MKL \texttt{DGEMM} and VML
  \texttt{VEXP}.
}
\label{tab:ks}
\end{table}  

All the algorithms described in this paper rely on multiplying
submatrices of $K$ with vectors, which we refer to as \emph{``kernel
  summation''}. These matrices can be precomputed and stored or they
can be used in a matrix-free manner, by computing $K_{ij} =
\Ker(\bx_i,\bx_j)$ in $\MA{O}(d)$ time on the fly.

For example, during the factorization of $K$, submatrices
$K_{\sk{\beta}\alpha}$ can be computed and stored.
In this case, MatVec of $K_{\sk{\beta}\alpha}$ for all treenodes $\alpha$ in the
solving phase can be done in $\MA{O}(sN\log N)$ with \texttt{GEMV}.
However, storing all these submatrices requires $\MA{O}(sN\log N)$ memory.
Memory requirements are even higher in the level-restricted version of
our algorithm. Thus, we never store these submatrices in the hybrid methods
due to the storage requirements.

Alternatively a matrix-free version only requires $\bigO(dN)$ storage (the coordinates of
the points) and turns the $\texttt{GEMV}$ to a
$\texttt{GEMM}$. However, since kernel evaluations are quite expensive,
this calculation can be significantly slower than storing the matrix
and computing summation using \texttt{GEMV}. Our goal is to reduce the
storage requirements without significantly sacrificing performance.

In~\cite{march-xiao-yu-biros15}, we presented
\texttt{GSKS} (General Stride Kernel Summation), an
matrix-free kernel summation that performs fusing optimization. 
While the best-known method computes
\begin{equation}
K_{\sk{\beta}\alpha}u =
  \texttt{GEMV}(\Ker(\texttt{GEMM}(\XX_{\sk{\beta}}^{T},\XX_{\alpha})),u),
\end{equation}
\texttt{GSKS} fuses $\Ker$ (kernel function) and \texttt{GEMV}
(reduction) into \texttt{GEMM} (semi-ring rank-$d$ update).
\cite{KNN:SC2015} uses the same idea to fuse nearest-neighbor search
into \texttt{GEMM}.  With a BLIS-like framework~\cite{van2015blis},
matrix-matrix multiplication ($C=AB$) is divided into subproblems.  A
small subproblem that fits $C$ into registers is implemented in
vectorized assembly or intrinsic to maximize FLOPS throughput.  The
idea is to directly perform kernel evaluation and the \texttt{GEMV} on
$C$ while it is still in the register and only store back a vector
$w$.  In short, for a typical kernel summation that involves an $m
\times n \times d$ \texttt{GEMM} with $\MA{O}(md+nd+mn)$ MOPS (Memory
Operations) in the best known method, \texttt{GSKS} can achieve
$\MA{O}(mnd)$ FLOPS but with only $\MA{O}(md+nd)$ MOPS.  This helps
the computation become less memory bound even with small $d$.  In this
work, we implement this idea in \texttt{AVX2} and \texttt{AVX512} for
Haswell and KNL architectures.  We present the performance of these
two different approaches in \tabref{tab:ks}.  Due to the $\MA{O}(mn)$
memory saving, \texttt{GSKS} is about $3\sim30$x faster than the best
known method on KNL for large problem size\footnote{\scriptsize{For
small problem size, \texttt{GEMM} in LIBSXMM~\url{https://github.com/hfp/libxsmm}
may be slightly faster than MKL \texttt{GEMM}, but it is still memory bound
when $d$ is small.}} and $d<68$.  We see that using \texttt{GSKS}
significantly outperforms using the standard approach.


\section{Theory} \label{s:theory} Here, we present some theoretical complexity guarantees and discuss the 
stability of our direct solver.

\textbf{Work.}
We present the complexity analysis of Algorithms \ref{a:factor}, \ref{a:solve}
and \ref{a:iterativesolve}.
Throughout, we fix the leaf size $\ppl$, level restriction $L$, and maximum 
skeleton size $\ns$.
$T^f(N)$ denotes the complexity of \algref{a:factor}, and $T^s(N)$ of 
\algref{a:solve}, each for $N$ points. Since {\tt Solve} does either an 
LU solve or matrix-vector multiply in each step, we have
\begin{equation}
  T^s(N) = 2T^s(N/2)+\MA{O}(Ns+s^2) = \MA{O}(sN\log{N}).
\end{equation}
Notice that solving $\hat{P}_{\alpha\sk{\alpha}} =
\sk{K}_{\alpha\alpha}^{-1}P_{\alpha\sk{\alpha}}$ does not require
traversing to the leaf level.
Instead \eqref{e:invtelescope} only takes $\MA{O}(s^2N)$ work.  Using
the complexity above, we derive $T^f(N)$ as
\begin{equation}
  T^f(N) = 2T^f(N/2)+s^2 N + s^3 = \MA{O}(s^2 N\log{N}).
\end{equation}
In the hybrid solver, each $(I+V W)x$ operation requires
$\bigO(2^{L}sN)$ work.  To summarize, both \algref{a:factor}
and \algref{a:solve} take $\MA{O}(N\log{N})$ work,
and \algref{a:iterativesolve} takes $\MA{O}(N\log{N})$ with additional
$\MA{O}(N)$ for each iteration if $L$ is independent from $N$.

\textbf{Communication.}
The communication cost for the solving phase is $\MA{O}(s\log p)$ per level.
To traverse the whole tree, $\MA{O}(s\log^2 p)$ is required per right hand
side.
However, during the factorization the solving phase does not recur. Thus,
  instead of $\MA{O}(s^2\log^2 p)$ for $s$ right hand sides, there is only
  $\MA{O}(s^2\log p)$ communication per level.
Overall, the communication cost for the full factorization is $\MA{O}(s^2\log^2 p)$
since there are $\log p$ distributed levels.

\textbf{Memory.}
The memory cost of our methods depend on level restriction $L$ and
maximum skeleton size $\ns$. 
These requirements are in addition to the cost to store the
coordinates and skeleton information for \ASKIT{}, reported
in \cite{march-xiao-biros15}.  In our direct solver, we require the
factors $U$, $V$, and $I + WV$ for each level of the tree below the
level-restriction $L$ in which skeletonization stops. This requires
$\bigO(2 \ns N + \ns^2)$ per level.
Therefore, the overall memory required for our method is
\begin{equation}
  \label{e:storage}
\bigO \left( \left(2 \ns N + \ns^2\right) \left(
      \log\left(\frac{N}{\ppl}\right) - L \right)  \right).
\end{equation}
Using \texttt{GSKS} can reduce $sN\log(N/m)$ to $\MA{O}(1)$ by computing $V$ on the fly. 
Recomputing $W$ with \eqref{e:invtelescope} can reduce another $sN\log(N/m)$ to
$sN$. Using both schemes yields
$\MA{O}(s^2(\log(N/m)-L)+sN)$ storage with $\MA{O}((d+s^2)N\log N)$ work
(still $\MA{O}(N\log N)$ asymptotically).


\textbf{Stability.}
Overall, the stability of our method is related to the conditioning of
$(\lambda I + \sk{K})$, $D$ and the reduced system $(I+VW)$. We use
$\kappa = \sigma_{1} /
\sigma_{\textrm{max}}$ to denote the 2-norm condition number of a matrix where $\sigma_{1}$ 
and $\sigma_{\textrm{max}}$ are the largest and smallest singular
values of the matrix.

\cite{hager1989updating} suggests that when either $U$ or $V$ are orthonormal,
then $\kappa(I+VW) \leq \kappa(D)\kappa(\sk{K})$. Although, our $U$ and $V$ are
not orthonormal, in our experience $\kappa(I+VW)$
does not have a conditioning problem when $D$ and $\sk{K}$ are well-conditioned.
The relation between $\kappa(\lambda I+D)$ and $\kappa(\lambda I + K)$
is more interesting.
%
In general when $h$ shrinks, we expect $K$ to become more
diagonally dominant and thus better conditioned.
However, counter to this intuition, it is possible for $D$ to become 
\emph{more} poorly conditioned as $h$ shrinks.

Since $D$ is a submatrix of $K$, we have
$\sigma_{1}(D) \leq \sigma_{1}(K)$ and
$\sigma_{n}(D) \leq \sigma_{n}(K)$. When $\sigma_{n}(K) < \lambda$, then 
$\kappa(\lambda I + D )  < \kappa(\lambda I + K)$ since $\lambda$
dominates in the denominator.
%
However when $\sigma_{n}(\sk{K}) > \lambda$, $\kappa(\lambda I + D)$
can grow even as $\kappa(\lambda I + K)$ remains small.  If this case
happens in many levels of our factorization, then the method is not
stable. With narrow bandwidths where $K$ approaches a
(blocked)-diagonal matrix, $\sigma_{n} > \lambda$ may occur.  


Under the framework of hierarchical matrices, the pivoting rows we
can choose during the $D$ factorization are limited to the skeleton
rows. Thus, even $\kappa(\lambda I + K)$ is not bad, $(\lambda I
+ D)$ can be unstable due to the aggressive pivoting strategy if
$\lambda$ is small. 
Our methods can detect this situation, but avoiding this case entirely
(or fixing it) is not straightforward. %
%



\section{Experimental Setup} \label{s:setup} We performed numerical experiments on Haswell and KNL architectures
with four different setups to examine the accuracy and efficiency of
our methods.  Especially, we want to demonstrate (1) the complexity
improvement against ~\cite{yu-march-xiao-biros16}, (2) FLOPS
efficiency, (3) scalability and (4) the advantages of our hybrid
solver.  We explore the task of kernel ridge regression for binary
supervised classification~\cite{wasserman04}, which requires
approximating the solution of $(\lambda I + K)^{-1}$ during the
training step. We use the Gaussian kernel with bandwidth $h$.  The
model weight $w$ is chosen by solving $w = (\lambda I
+ \sk{K})^{-1}u$, where $u$ is given (the labels).  Once $w$ is
computed, the label given by $\bx \notin \XX$ is
$\text{sign}(\Ker(\bx,\XX)\bw)$.  We apply our methods to train this
model on real-world datasets employing up to 3,072 x86 cores and 4,352
KNL cores. The the percentage of correct predictions (Acc) is reported
in \tabref{tab:dataset}, along with the optimal $h$ and $\lambda$ that
were found using holdout cross validation.

\textbf{Implementation and hardware.} 
Our experiments were conducted on
Lonestar5 (two 12-core, 2.6GHz, Xeon E5-2690 v3  ``Haswell''
per node) and Stampede (68-core, 1.4GHz, Xeon Phi 7250 ``KNL'' per node) clusters at the Texas Advanced Computing Center. 
The theoretical peak\footnote{\scriptsize{We estimate the peak according to the clockrate
and the theoretical \texttt{FMA} throughput. For 24 Haswell cores,
$998=2\times12\times2.6\times16$. For 68 KNL cores,
$3046=68\times1.4\times32$.
As a reference, MKL \texttt{GEMM} can achieve $87\%$ on the Haswell node and 
$69\%$ on the KNL node.
We assume two VPUs can dual issue \texttt{DFMA}s~\cite{sodani2016knights}.
However, Intel processors may have a different frequency while fully issuing 
  \texttt{FMA}, and the clockrate may drop to 1.0 GHz. This may be the reason
why MKL \texttt{GEMM} can only achieve 2.1 TFLOPS on KNL.}}
performance is 998 GFLOPS per Haswell node and 3,046 GFLOPS
per KNL node.
Inv-Askit and \texttt{GSKS} are compiled with \texttt{intel-16 -O3 -mavx} on
Lonestar5 and \texttt{intel-17 -O3 -xMIC-AVX512} on Stampede.
All iterative solvers employ a
Krylov subspace method (\GMRES{}) from the PETSc
library~\cite{petsc-web-page}.  Specifically, we use modified
Gram-Schmidt for re-orthogonalization and employ \GMRES{} CGS
refinement.
If not specified, KNL experiments use Cache-Quadrant configuration with
$\texttt{OMP\_PROC\_BIND=spread}$.
``\rm{T}'' refers to the total runtime in seconds, and ``\rm{GFs}'' refers to the
\rm{GFLOPS} per node.


\begin{table}[!t]
\centering
{ \begin{tabular}{|r|rrrr|r|} 
  \hline 
  \rowcolor[gray]{0.8}
	Dataset & $N$ & $d$ & $h$ & $\lambda$ & Acc \\
  \hline
  \cellcolor[gray]{0.8} \textbf{COVTYPE} & 0.1--0.5M & 54 & .07 & .3 & 96\% \\ 
  \cellcolor[gray]{0.8} \textbf{SUSY}    & 4.5M &  8 & .07 & 10 & 78\%  \\ 
  \cellcolor[gray]{0.8} \textbf{MNIST2M} & 1.6M & 784 & .30 & 0 & 100\% \\ 
  \cellcolor[gray]{0.8} \textbf{HIGGS}   & 10.5M & 28 & .90 & .01 & 73\% \\
  \hline
  \cellcolor[gray]{0.8} \textbf{MRI} & 3.2M & 128 & 3.5 & 10 & - \\ 
  \cellcolor[gray]{0.8} \textbf{MNIST8M} & 8.1M & 784 & 1.0 & 1.0 & - \\ 
  \cellcolor[gray]{0.8} \textbf{NORMAL}  & 1--32M   & 64 & .19 & 1.0 & - \\ 
  \hline
  \end{tabular}
  } \caption{Datasets used in the experiments. Here
  $N$ denotes the size of the training set, and $d$ is the dimensionality of 
	points in the dataset.
  The testing sets are disjoint from the training sets. We sample 10K testing
  points and report the binary classification accuracy in the ``Acc'' column.
  $h$ is the bandwidth of the Gaussian kernel used in our
  experiments.
  The regression results are produced using the parameters above. 
  Some combinations we used during the cross-validation are presented in 
  details in \secref{s:results}.
  \textbf{MRI}, \textbf{MNIST8M} and \textbf{NORMAL} are not used in regression
  tasks. For the \textbf{MNIST2M} we perform one-vs-all binary
  classification for the digit '3'. All coordinates are normalized to
  have zero mean and unit variance. 
  }
\label{tab:dataset}
\end{table}

\textbf{Datasets.} We use real-world datasets: 
\textbf{COVTYPE} (forest cartographic variables); \textbf{SUSY}  and
\textbf{HIGGS} (high-energy physics)~\cite{Lichman:2013};
\textbf{MNIST} (handwritten digit recognition)~\cite{chang2011libsvm}; and \textbf{MRI} (brain MRI)~\cite{menze:hal-00935640}.
We also use a 64D synthetic dataset, which is drawn from a 6D Normal
distribution and embedded in 64D with additional noise.  This set is a
dataset with a high ambient but relatively small intrinsic dimension.

\textbf{Accuracy metrics and parameter selection.}
For the linear solve, we report the relative residual 
\begin{equation}
  \label{e:errors}
  \begin{split}
  \epsilon_{r} & = \|u - (\lambda I + \tilde{K})w\|_2 / \|u\|_2.
  \end{split}
\end{equation}
The parameters $h$ and $\lambda$ used in the Gaussian kernel were
selected using cross-validation. 
In \tabref{tab:dataset} we report the
parameters we used.  Other combinations in \secref{s:results} are
candidates for the cross-validation. 
Level restriction $L$ is choosen such that the relative error is
controlled. In \tabref{tab:knl_single_node}, \ref{tab:performance} 
we use $L=3$, and for experiments in \figref{fig:converge} we use
$L=5$ or $7$.




\section{Empirical Results} \label{s:results} 
\begin{figure}[!t]
  \centering
  \includegraphics[scale=.285]{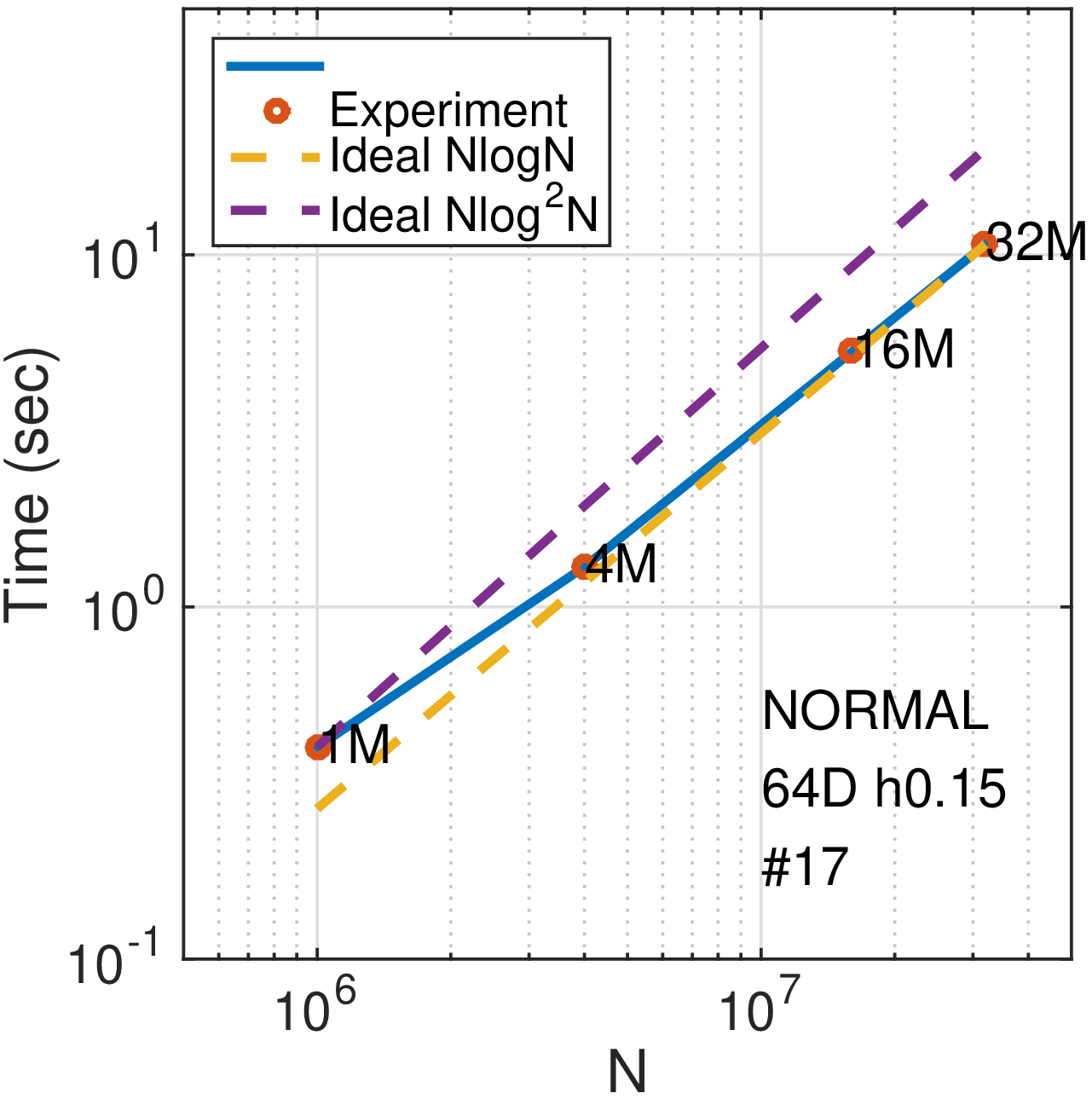}
  \includegraphics[scale=.285]{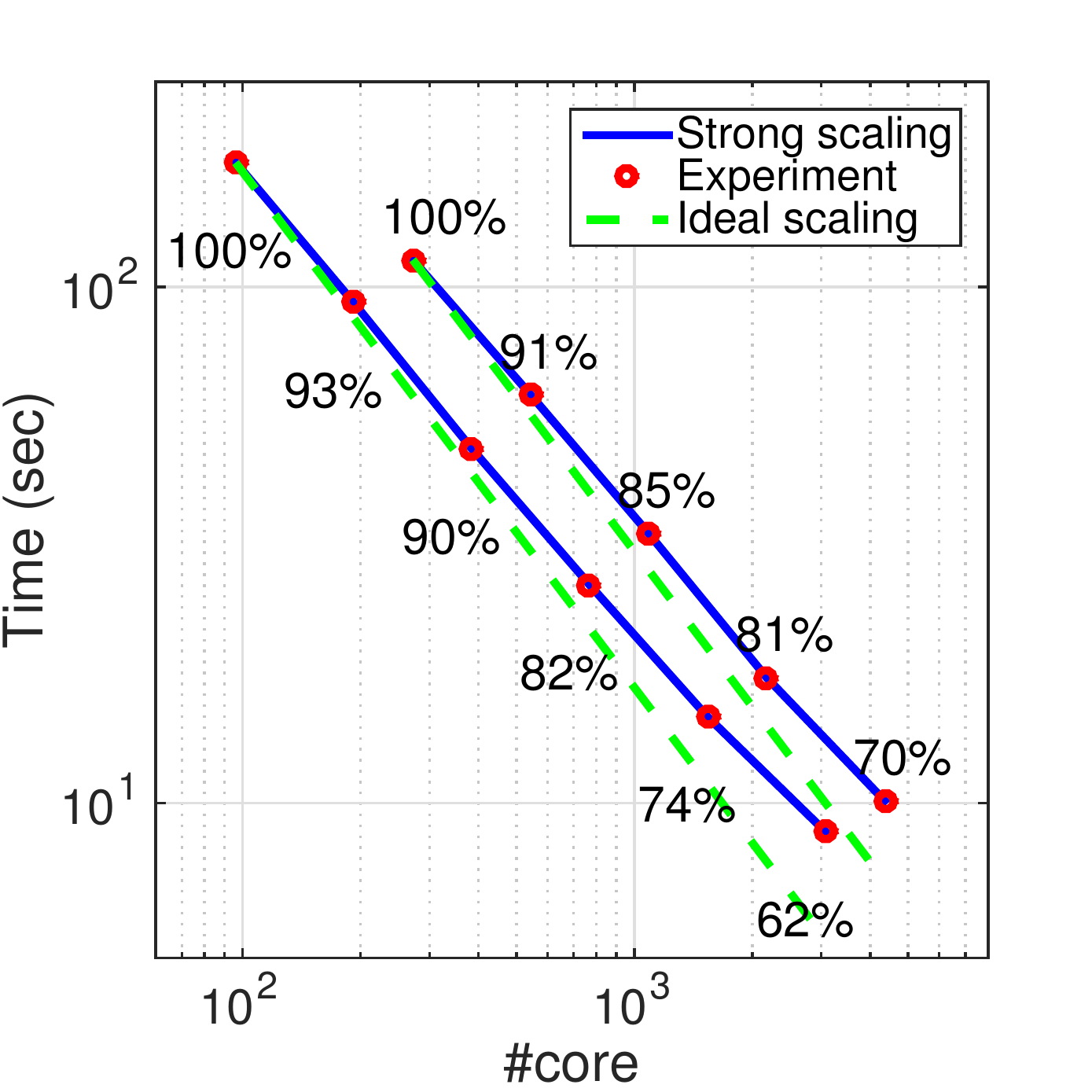}
  \caption{$\MA{O}(N \log N)$ verification (\#\ref{exp:loglinear}) and strong scaling
    (\#\ref{exp:strong}).
    \#\ref{exp:loglinear} use \textbf{NORMAL} with $m=512$, $\kappa=128$, a fixed
    $s=256$ and $L=1$.
	The blue lines are experimental factorization time, and
	yellow lines are theoretical (ideal) time.
  \#\ref{exp:strong} use \textbf{NORMAL1M} with $\kappa=128$, $m=s=2048$ and $L=1$.
  We increase number of nodes up to 128 Haswell nodes (3,072 cores) and 64 KNL
  nodes (4,352 cores). The green lines represent the ideal scaling.
  }
  \label{fig:scaling}
\end{figure}

\begin{figure}[!t]
  \centering
  \includegraphics[scale=.42]{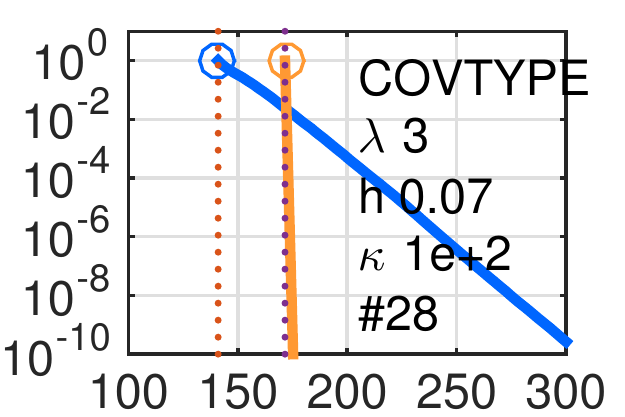}
  \includegraphics[scale=.42]{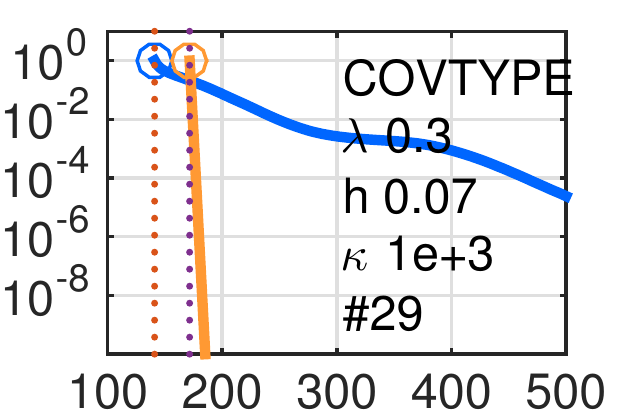}
  \includegraphics[scale=.42]{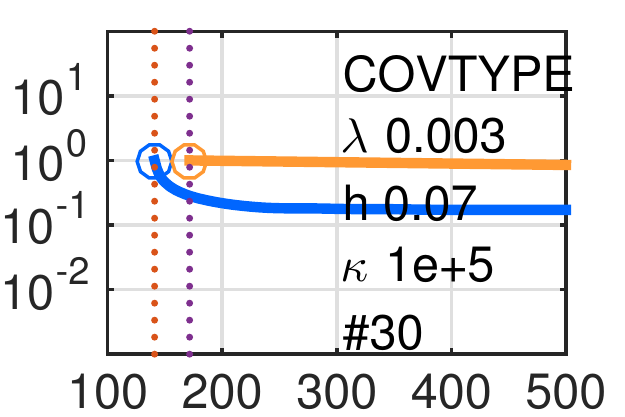}
  \includegraphics[scale=.42]{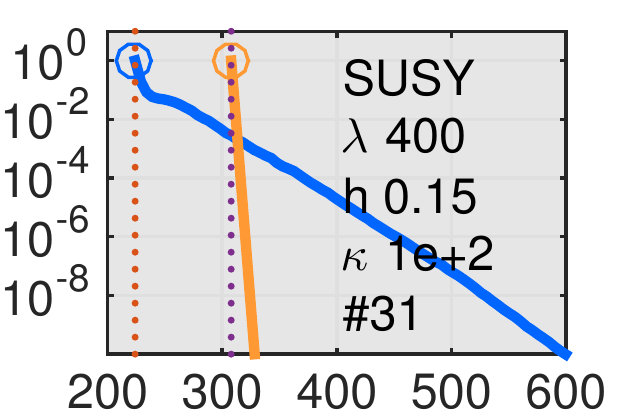}
  \includegraphics[scale=.42]{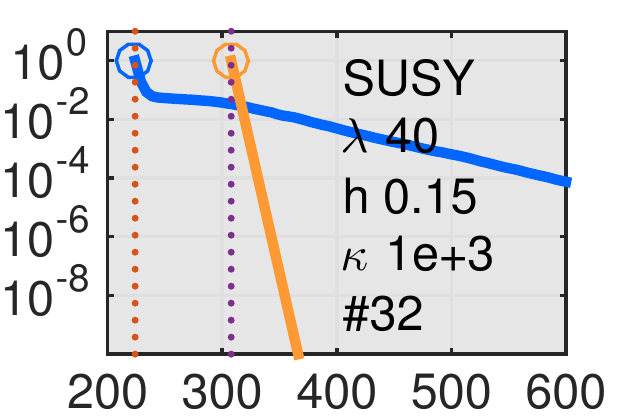}
  \includegraphics[scale=.42]{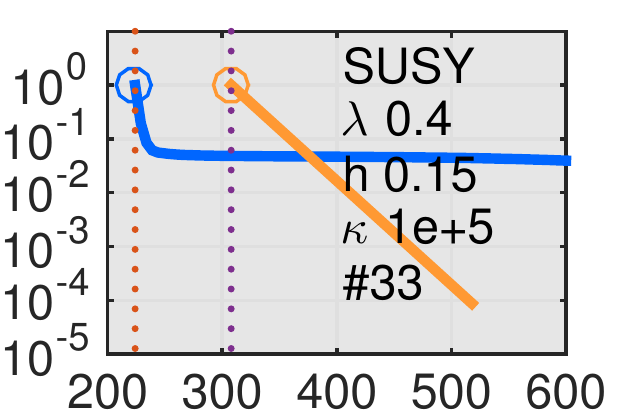}
  \includegraphics[scale=.42]{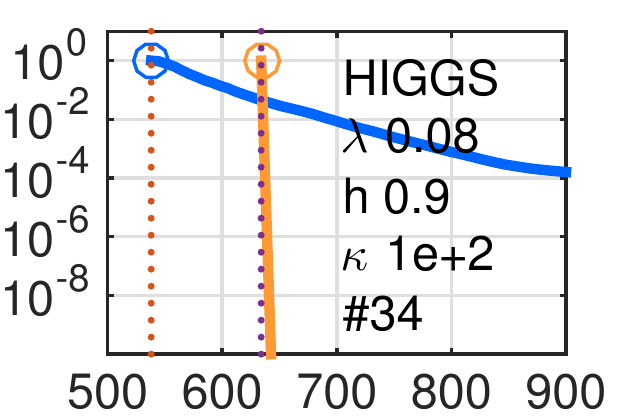}
  \includegraphics[scale=.42]{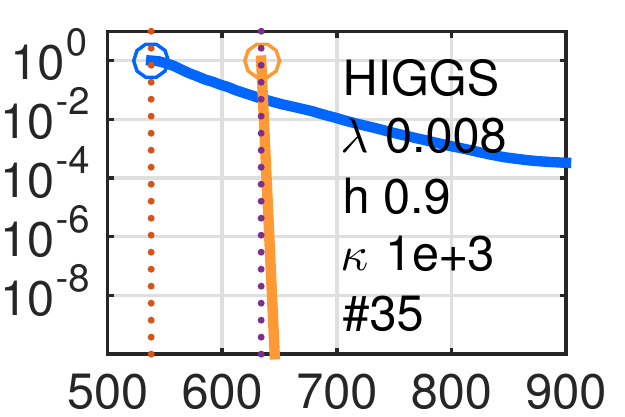}
  \includegraphics[scale=.42]{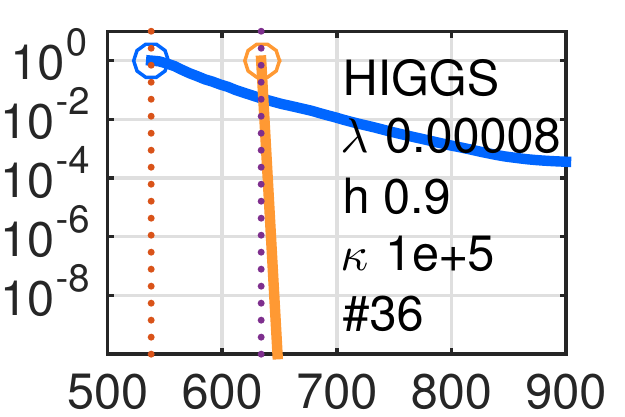}
  \includegraphics[scale=.42]{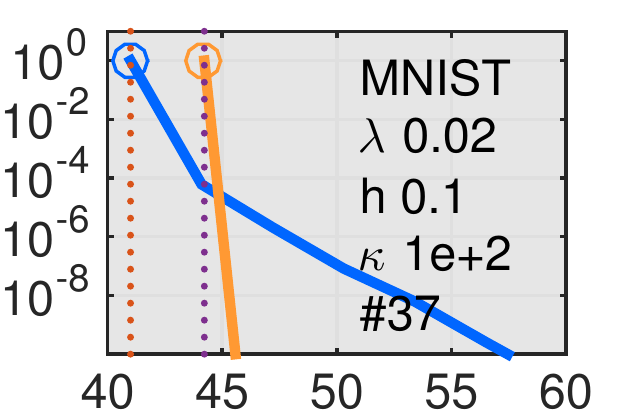}
  \includegraphics[scale=.42]{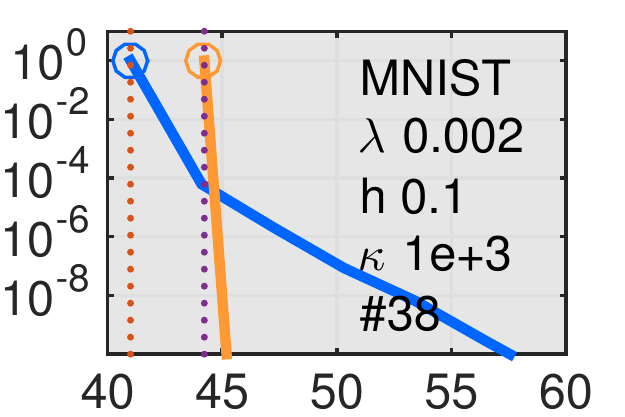}
  \includegraphics[scale=.42]{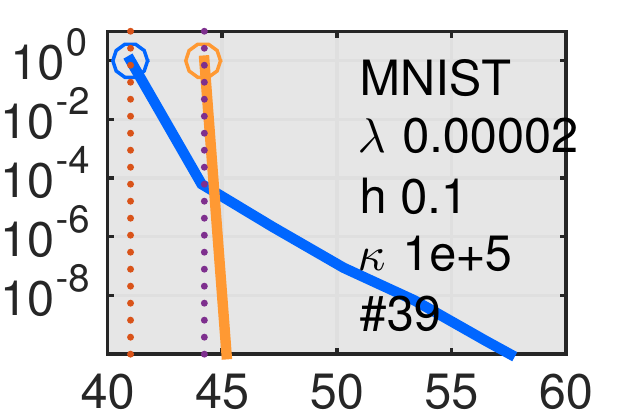}
  \caption{solving $\lambda I + \sk{K}$: Convergence  of the relative residual $\epsilon_{r}$  
  (vertical axis) over time (horizontal axis) in seconds.
  \textbf{(a)}  (blue) is unpreconditioned GMRES;
  \textbf{(b)} (orange) is hybrid method.	All experiments used $\tau=1E-5$.
	$m$, $k$ and $s_{\rm{max}}$ are the same as \tabref{tab:speedup}; $\kappa$ is the condition number.
    \textbf{COVTYPE} and \textbf{HIGGS} used $L=5$ restriction.
	\textbf{SUSY} and \textbf{MNIST} used $L=7$. 
  These runs may be 100--1000$\times$ more expensive (also
  running out of memory) if we were to use the direct
  solver. } \label{fig:converge}
\end{figure}

The experiments are labeled $\#1$ to $\#39$ in the tables and figures.
We select representative parameter combinations and compare the
runtime between \cite{yu-march-xiao-biros16} and our 
\algref{a:distributedfactor} in \tabref{tab:speedup}. 
We present single node performance on Haswell and KNL 
in \tabref{tab:knl_single_node}.
In~\figref{fig:scaling}, we present strong scalability and verify the $\MA{O}(N
\log N)$
complexity of our methods (\algref{a:distributedfactor}).
In \tabref{tab:performance}, we compare our hybrid
(\algref{a:iterativesolve}) with the direct method
(\algref{a:distributedsolve}).  In \figref{fig:converge}, we report
the convergence behavior of iterative solver
(\algref{a:iterativesolve}) on $\lambda I + \sk{K}$ and our hybrid
factorization.  Here the parameter $\tau$ indicates the relative
tolerance of approximation of the kernel matrix $K$, $\nsmax$ is the
maximum skeleton size, $\nk$ is the number of nearest neighbors used
for skeletonization sampling in \ASKIT{}, $\ppl$ is the leaf node
size, and $L$ is the level restriction.

\begin{table}[!t]
\setlength\tabcolsep{4.8pt}
\centering
{
  \begin{tabular}{|r|>{\columncolor[gray]{0.8}}r|r|rr|rr|rr|} 
  \hline
  \rowcolor[gray]{0.8}
  \multicolumn{3}{|c|}{$\tau$} & \multicolumn{2}{c|}{1E-1} &
	\multicolumn{2}{c|}{1E-3} & \multicolumn{2}{c|}{1E-5}  \\
  \hline
  \rowcolor[gray]{0.8}
  \# &dataset& $h$ & $\log^2$ & $\log$ & $\log^2$ & $\log$ & $\log^2$ & $\log$ \\
  \hline
  \rownumber\label{exp:covtypetime35} &\textbf{COVTYPE}& .35 & $<1$ & $<1$ & 5 & 3 & 15 & 8 \\ 
  \rownumber\label{exp:covtypetime07} & & .07 & 5 & 3 & 24 & 11 & 27 & 12 \\ 
  \hline
  \rownumber\label{exp:susytime50} &\textbf{SUSY}& .50 & $<1$ & $<1$ & 1 & $<1$ & 6 & 3 \\ 
  \rownumber\label{exp:susytime05} & & .05 & 63 & 23 & 125 & 37 & 125 & 37 \\ 
  \hline
  \rownumber &\textbf{MNIST2M}& 1.0 & 24 & 11 & 38 & 15 & 40 & 16 \\ 
  \rownumber & & .10 & 1 & $<1$ & 1 & $<1$ & 4 & 7 \\ 
  \hline
  \rownumber &\textbf{MNIST8M}& 0.3 & 142 & 51 & 185 & 61 & 195 & 64 \\ 
  \hline
  \rownumber\label{exp:higgstime20} &\textbf{HIGGS}& 2.0 & 75 & 29 & 324 & 84 & 326 & 85 \\ 
  \rownumber\label{exp:higgstime09} && .90 & 216 & 66 & 317 & 83 & 323 & 85 \\ 
  \hline
  \rownumber\label{exp:last} &\textbf{NORMAL32M}& .19 & 18 & 10 & 46 & 22 & 84 & 28 \\ 
  \hline
  \end{tabular}
}
\caption{Factorization time comparison in second. Experiments are done
  on Lonestar5 using 128 nodes (3,072 cores) with adaptive ranks $s$ selected 
  by $\tau$ and $s_{\rm{max}}$. 
    \textbf{COVTYPE} used $m=2,048$, $\kappa=2,048$, $s_{\rm{max}}=2,048$.
	\textbf{SUSY} used the same combination.
	\textbf{MNIST} used $\kappa=256$.
	\textbf{HIGGS} used $m=512$, $\kappa=1,024$.
	\textbf{NORMAL} used $m=512$, $\kappa=128$ and $s_{\rm{max}}=256$.
}
\label{tab:speedup}
\end{table}  

\textbf{Comparison with \cite{yu-march-xiao-biros16} (\tabref{tab:speedup}).} 
We compare the factorization time between the $\MA{O}(N\log^{2}N)$ algorithm
\cite{yu-march-xiao-biros16} and our $\MA{O}(N \log N)$ algorithms using the
same parameters. We only 
compare the case without level restriction,
because \cite{yu-march-xiao-biros16} does not support this feature. 
The runtime is directly associated with the rank $\ns$.  For example,
the $U$, $V$ matrices in \#\ref{exp:susytime05} are much larger than
those in \#\ref{exp:susytime50}. Thus, the runtime is also much longer.
The overall speedup is about 2--4$\times$ due to the $\log{N}$ term. Both
methods construct exactly the same factorization (up to roundoff
errors).  Although the speedup is not exactly $\log{N}$ due to the
prefactors depending on $s$ and $d$, we can expect the asymptotic
speedup to be $O(\log{N})$.  For example, \textbf{COVTYPE} can only
achieve $1.9\times$, but
\textbf{HIGGS} can achieve $3.8\times$ because the problem size
is $20\times$ larger.
Both methods have $N \log N$ 
complexity for the ``Solve'' operation. For the experiments in
\tabref{tab:speedup}, the longest ``Solve'' 
operation (\#\ref{exp:higgstime20}) is less than 2 seconds.  

\begin{table}[!t]
\centering
{
  \begin{tabular}{|>{\columncolor[gray]{0.8}}r|rr|rr|r|r|}
  \hline
  \# & 
    \rownumber\label{exp:has1} & \rownumber\label{exp:has4} &
    \rownumber\label{exp:cq1} & \rownumber\label{exp:cq4} &
    \rownumber\label{exp:fq1} & \rownumber\label{exp:fsnc4} \\
  \hline
  \rowcolor[gray]{0.8}
  Config & \multicolumn{2}{c}{Haswell} & \multicolumn{2}{c|}{Cache-Quad} & F-Quad & F-SNC4 \\
  \hline
  p    &  1 & 4 & 1 & 4 & 1 & 4 \\
  nthd & 24 & 6 & 68 & 17 & 68 & 17 \\
  \hline
  $\rm{T_{f}}$ & 89 & 94 & 41  & 52 & 62 & 77 \\
  $\rm{GF_{f}}$ & 623 & 592 & 1356 & 1136 & 895 & 723 \\
  \hline
  \rowcolor[gray]{0.8}
  \multicolumn{7}{|c|}{Compute MatVec $V$ with \texttt{GEMV}.} \\
  \hline
  $\rm{T_{s}}$ & 0.8 & 0.8 & 0.9 & 0.9 & 0.9 & 1.0 \\
  $\rm{GF_{s}}$ & 14 &  14 & 12  & 12 & 12 & 11 \\
  \hline
  \rowcolor[gray]{0.8}
  \multicolumn{7}{|c|}{Reevaluate $V$ with \texttt{GEMM}.} \\
  \hline
  $\rm{T_{s}}$ &  4.4 & 5.3 & 7.4 & 4.0 & 7.5 & 5.8 \\
  $\rm{GF_{s}}$ & 158 & 119 &  40 &  74 &  39 &  51 \\
  \hline
  \rowcolor[gray]{0.8}
  \multicolumn{7}{|c|}{Compute MatVec $V$ with \texttt{GSKS}.} \\
  \hline
  $\rm{T_{s}}$  & 1.1 & 1.8 & 1.1 & 1.2 & 1.2 & 1.3 \\
  $\rm{GF_{s}}$ & 269 & 164 & 269 & 243 & 243 & 228 \\
  \hline
  \end{tabular}
}
\caption{Single node performance on \textbf{COVTYPE100K} with $m=s=2048$ 
  (fixed rank) and $L=3$.
  $p$ is the number of MPI processes, and $nthd$ is the number of OpenMP
  threads per process.
  \#\ref{exp:has1} and \#\ref{exp:has4} were conducted on Haswell,
  and \#\ref{exp:cq1} to \#\ref{exp:fsnc4} where conducted on KNL
  with different configurations.
  \#\ref{exp:cq1} and \#\ref{exp:cq4} use cache-quadrant. \#\ref{exp:fq1}
  uses flat-quadrant, and \#\ref{exp:fsnc4} uses flat-snc-4 configuration.
  \#\ref{exp:has1} achieves the high factorization performance on Haswell,
  and \#\ref{exp:cq1} is the most efficient configuration on KNL.
  In each column, we report factorization time ($\rm{T_{f}}$) and GFLOPS ($\rm{GF_{f}}$) and three 
  different solving time ($\rm{T_{s}}$). These three solving schemes have different
  storage requirement and memory operations.
  }
\label{tab:knl_single_node}
\end{table}

\textbf{Single node performance (\tabref{tab:knl_single_node}).}
We conduct a set of single node experiments to show the FLOP rates we
achieve and to test some of the memory models on KNL.  On a Haswell
node, \#\ref{exp:has1} reaches $62\%$ $(623/998)$ of the
theoretical peak.  For a KNL node, \#\ref{exp:cq1} (Cache-Quadrant) is
the fastest and achieves $45\%$ $(1356/3046)$.  Using MPI on KNL ($p>1$) is typically slower
due to the extra memory operations.
%
%
Our implementation does not perform very well on the flat memory mode
(\#\ref{exp:fq1} and \ref{exp:fsnc4}).  The memory requirements
usually exceed 16GB and as a result $U$ and $V$ cannot fit into
MCDRAM. We tried manually swapping memory between MCDRAM and DDR4 but
this was not as efficient as using the cache memory mode.
%

\textbf{Reducing storage (\tabref{tab:knl_single_node}).} 
In \tabref{tab:knl_single_node}, we report three different  schemes for kernel summation~\secref{s:ks}: \texttt{GEMV},
\texttt{GEMM} and \texttt{GSKS}.  
The first scheme takes $\MA{O}(sN\log{N})$ time and space.  The last
two schemes evaluate $K_{\sk{\beta}\alpha}u$ in $\MA{O}(dsN\log{N})$
where $d$ is $54$ in this dataset.
\texttt{GEMM} takes $\MA{O}(sN)$ to store the matrix,
but \texttt{GSKS} is matrix free with $\MA{O}(1)$ space.
\texttt{GSKS} is only $1.2-1.6\times$ slower than the
\texttt{GEMV} approach while it is $4-7\times$ faster than the \texttt{GEMM}
approach.  Notice that 80\% of \rm{$T_{s}$} is dominated
by \texttt{GETRS} (triangular solver) when $d$ is
small. Since \texttt{GETRS} can only achieve 2 GFLOPS on a KNL node,
the overall \rm{$GF_{s}$} is somewhat low.



\textbf{Scaling (\figref{fig:scaling}).}
In \#\rownumber\label{exp:loglinear}, we use 128 nodes (3,072 cores)
and increase $N$ from 1M to 32M.  We can observe that our
implementation is very close to the theoretical $N \log N$ scaling
(yellow) but lower than the $N \log^2 N$ scaling (purple).
In \#\rownumber\label{exp:strong}, we fix the data set (\textbf{NORMAL
1M}) and increase the number of cores. The green line is the ideal
scaling ($100\%$), and our implementation reaches $62\%$ efficiency on
3,072 Haswell cores and $70\%$ on 4,352 KNL cores.  This relatively
small problem (1M) cannot fully exploit all computing resources; thus,
we can see the degradation while $N$ is small or when the number of
cores is large ($\sim230$ points per core for 64 KNL nodes).

\begin{table}[!t]
\setlength\tabcolsep{5.2pt}
\centering
{
  \begin{tabular}{|r|>{\columncolor[gray]{0.8}}r|r|rr|rr|r|r|}
  \hline
  \rowcolor[gray]{0.8}
  \# & KNL & \ASKIT{} & $\rm{T_{f}}$ & $\rm{GF_{f}}$ & $\rm{T_{s}}$ & $\rm{GF_{s}}$ & $\epsilon_{r}$ & KSP \\
  \hline
  \multicolumn{9}{|c|}{\bf{SUSY $h=0.15$ $\lambda=40$}} \\
  \hline
  \rownumber\label{exp:susyknl} & x & 294 & 60 & 844 & 1.5 & 82 & 5e-12 & - \\
  \rownumber\label{exp:susyhas} & - & 570 & 110 & 467 & 1.2 & 99 & 5e-12 & - \\
  \color{orange}
  \rownumber\label{exp:susyksp} & - & 544 & 59 & 414 & 22.3 & 246 & 6e-4 & 98 \\
  \hline
  \multicolumn{9}{|c|}{\bf{MRI $h=3.5$ $\lambda=10$}} \\
  \hline
  \rownumber\label{exp:mriknl} & x & 302 & 46 & 795 & 1.4 & 319 & 1e-10 & - \\
  \rownumber\label{exp:mrihas} & - & 396 & 84 & 427 & 1.5 & 298  & 1e-10 & - \\
  \color{orange}
  \rownumber\label{exp:mriksp} & - & 217 & 37 & 467 & 39.6 & 508 & 3e-4 & 93 \\
  \hline
  \multicolumn{9}{|c|}{\bf{MNIST2M $h=1.0$ $\lambda=1$}} \\
  \hline
  \rownumber\label{exp:mnistknl} & x & 237 & 27 & 655 & 1.9 & 592 & 1e-13 & - \\
  \rownumber\label{exp:mnisthas} & - & 270 & 47  & 372 & 2.3 & 489 & 1e-13 & - \\
  \color{orange}
  \rownumber\label{exp:mnistksp} & - & 217 & 19  & 404 & 27.2 & 612 & 1e-3 & 27 \\
  \hline
  \end{tabular}
}
\caption{Hybrid and direct methods comparison withe level restriction $L=3$.
  All experiments use adaptive ranks with tolerance $\tau=0.00001$ and
  $s_{\rm{max}} = 2048$.  We report \ASKIT{} building time,
  factorization time ($\rm{T_{f}}$), solving time
  employing \texttt{GSKS} ($\rm{T_{s}}$) and efficiency in GFLOPS. The
  three experiments with orange index are the hybrid scheme and the
  remaining are the direct factorization.  We report the relative
  residual $\epsilon_{r}$ and number of Krylov iterations (KSP) (for the
  hybrid method).}
\label{tab:performance}
\end{table}  

\textbf{Hybrid and direct methods comparison (\tabref{tab:performance}).}
In \tabref{tab:performance} we set $L=3$ (level restriction) and
compare~\algref{a:iterativesolve} (hybrid) and~\algref{a:factor}
(direct) for problems that \algref{a:factor} can be applied (i.e., the full
factorization requires
$2^{L} \ns N + 2^{2L}s^2$ memory for level restriction $L$), with adaptive $\ns$ selection.
\#\ref{exp:susyhas}, \#\ref{exp:mrihas} and \#\ref{exp:mnisthas}  use the direct
factorization on Haswell, and \#\ref{exp:susyknl}, \#\ref{exp:mriknl},
and \#\ref{exp:mnistknl} on KNL. The remaining three runs are done on
Haswell using the hybrid method. On Haswell, we observe that the
factorization time $\rm{T_{f}}$ is about two times longer
than \#\ref{exp:susyksp}, \#\ref{exp:mriksp} and \#\ref{exp:mnistksp}.
In the factorization phase, both Haswell and KNL do not perform as
well as in~\tabref{tab:knl_single_node}, because using adaptive ranks $\ns$
results in load imbalance.  If we further increase $L$, as we need to do in 
\figref{fig:converge}, the cost of the full factorization can be
1000$\times$ in runtime and 30$\times$ in storage.

%

Applying the hybrid solver to a vector is slower than applying the
direct solver, due to the need of iteration.  E.g., $\rm{T_{s}}$
in \#\ref{exp:susyksp} is about 20$\times$ slower. Yet the overall
runtime ($\rm{T_{f}}+\rm{T_{s}}$) of the hybrid method is still
smaller than the direct one.  When $L$ is larger, the advantage of the
hybrid solver will be higher. For example, in~\figref{fig:converge}, we
report results that require $L=7$. \algref{a:factor} cannot be used:
the memory just for $Z$ with $\ns=2048$ exceeds 500GB.


\textbf{Convergence behavior for solving $\lambda I + \sk{K}$ (\figref{fig:converge}).}
We report the convergence rate
using four different bandwidths with two different
methods: \textbf{(a)} unpreconditioned GMRES using \ASKIT{}'s MatVec for $\lambda I
+ \sk{K}$ \textit{\textbf{\color{blue}(blue line)}} and 
\textbf{(b)} our hybrid method \algref{a:iterativesolve} \textit{\textbf{\color{orange}(orange line)}}. 
Each row corresponds to a dataset with a specific $h$.  These
experiments resemble a cross-validation study in which we vary
$\lambda$ in order to improve learning.
Across columns, we vary $\lambda$ as
$[10^{-2},10^{-3},10^{-5}]\sigma_1(\sk{K})$, where $\sigma_1(\sk{K})$
is an estimate, so that the condition number $\kappa$ of $\lambda I
+ \sk{K}$ is $10^2$, $10^3$ and $10^5$ respectively.  We report the
relative (to a zero initial guess) Krylov residual $\epsilon_{r}$
(y-axis) over time (x-axis).
The steeper the curve is, the faster the method converges.  The x-axis
offset represents setup costs.  E.g., in \#28, the offset of
the blue line ($\approx 140$ sec) is the cost of building the
tree and the skeletons (spent in \ASKIT{}).  The fixed cost
of \textbf{(b)} includes the fixed cost of \textbf{(a)} plus the
factorization time.


We see that most of the blue lines are flat when the condition number
is around 1E+5, but orange lines still decrease steadily except
for \#30 (see \secref{s:theory} for the stability issue). \emph{We can
observe 10--1000$\times$ speedup on the ``Solve'' operations.} Overall
the hybrid scheme is faster and has more predictable behavior.
\#30 is detected numerically ill-conditioning of $D$ in our solver.
Also notice that in \#30  both methods fail to converge.



\section{Conclusions} \label{s:conclusion} We have introduced new algorithms for approximately factorizing kernel
matrices.  We evaluated our algorithms on both real-world and
synthetic datasets with different parameters. We conducted analysis
and experiments to study the complexity and the scalability of our
methods.  These experiments include scaling up to 3,072 Haswell cores
and 4,352 KNL and exhibit significant speedups over existing
methods. The factorization can be very fast.  For example, it only
takes 10 seconds to factorize a kernel matrix with 32M points in 64D.
%
Our future work will focus on further optimization of our
implementation. In particular, we would like to introduce task
parallelism in the tree traversal to address the load balancing issue.
While adaptive ranks or adaptive level restriction is used, each
treenode may have different workload. In this case, scheduling is
important to avoid the critical path.
Additionally, we plan to address the stability issues mentioned
in \secref{s:theory} and explore other possible variants
(\emph{e.g.}~sparse off-diagonal blocks).



\newpage
\bibliographystyle{siam}
\bibliography{gb,refs}

\end{document}